\documentclass[aps,preprintnumbers,twocolumn,amsmath,amssymb,prb,showpacs]{revtex4}
\usepackage{bbm}
\usepackage{mathrsfs}
\usepackage{amsmath}
\usepackage{graphicx}
\usepackage{ulem}
\usepackage{color}
\usepackage{array}

\newcommand{\be}{\begin{equation}}
\newcommand{\ee}{\end{equation}}
\newcommand{\bea}{\begin{eqnarray}}
\newcommand{\eea}{\end{eqnarray}}
\newcommand{\bes}{\begin{split}}
\newcommand{\ees}{\end{split}}

\newcommand{\tr}{\operatorname{Tr}}

\newcommand{\bs}{\boldsymbol}

\begin{document}
\title{Deep learning of topological phase transitions from entanglement aspects: An unsupervised way}
\author{Yuan-Hong Tsai$^{1,2}$}\email{yhong.tsai@gmail.com}
\author{Kuo-Feng Chiu$^{3}$}
\author{Yong-Cheng Lai$^{3}$}
\author{Kuan-Jung Su$^{3}$}
\author{Tzu-Pei Yang$^{3}$}
\author{Tsung-Pao Cheng$^{3}$}
\author{Guang-Yu Huang$^{3}$}
\author{Ming-Chiang Chung$^{3,4,5}$}\email{mingchiangha@nchu.edu.tw} 
\affiliation{$^1$ AI Foundation, Taipei, 106, Taiwan}
\affiliation{$^2$ Taiwan AI Academy, New Taipei, 241, Taiwan}
\affiliation{$^3$ Physics Department, National Chung-Hsing University, Taichung, 40227, Taiwan}
\affiliation{$^4$ Physics Department, National Center for Theoretical Sciences, Taipei, 10617, Taiwan}
\affiliation{$^5$ Physics Department, Northeastern university, 360 Huntington Ave., Boston, Massachusetts 02115, U.S.A. }

\begin{abstract}
Machine learning techniques have been shown to be effective to recognize different phases of matter and produce phase diagrams in the parameter space interested, while they usually require prior labeled data to perform well. Here, we propose a machine learning procedure, mainly in an unsupervised manner, which can first identify topological/non-topological phases and then refine the locations of phase boundaries. By following this proposed procedure, we expand our previous work on the one-dimensional $p$-wave superconductor [Phys. Rev. B 102, 054512 (2020)] and further on the Su-Schrieffer-Heeger model, with an emphasis on using the quantum entanglement-based quantities as the input features. We find that our method not only reproduces similar results to the previous work with sharp phase boundaries but importantly it also does not rely on prior knowledge of the phase space, {\it e.g.}, the number of phases present. We conclude with a few remarks about its potential, limitations, and explainabilities.
\end{abstract} 

\date{\today}
\maketitle 

\section{Introduction}
Machine learning (ML) is not only a rapidly growing field of computer science recently with various applications from machine vision to natural language processing \cite{lecun15}, but also attracts much attention among researchers in the physical society. This technique is completely data-driven and thus a bottom-up method: Given a large amount of data or features, some function (often a neural network) is trained to correlate them with a more accessible form or condensed representations, which could be simply class labels or patterns. If such a trained function is well-generalizable, it can then predict (represent) an unknown new data point. Within several conceptual or practical applications in condensed matter physics, using ML to identify different phases of matter or to determine the phase boundaries are of particular interests \cite{Ohtsuki16,Nieuwenburg17,Carrasquilla17,Broecker17,Wetzel17,wang16,Tanaka17,Hu17,Broecker17b,Wetzel17b,Chng18,Liu18,Durr19,Kottmann20}. Moreover, it is quite remarkable that ML has also been shown insightful and to have great potential in considering topological phase transitions \cite{kim17a,kim17b,Zhang18,Sun18,Carvalho18,Ming19,Caio19,Zhang21,Scheurer19,Greplova20,Scheurer20},  where {\it no} obvious local order parameters are available.

There are two main approaches when using ML to classify phases of matter, no matter topological or symmetry-breaking. One is based on supervised learning, in which each training data sample should be labeled by a well-known regime (phase)  \cite{Ohtsuki16,Nieuwenburg17,Carrasquilla17,Broecker17,Wetzel17,kim17a,kim17b,Zhang18,Sun18,Carvalho18,Ming19,Caio19,Zhang21}. This approach often results in a better phase boundary decision while it requires prior knowledge of the underlying phases of the system such as the number of total phases in a focused parameter space. Therefore, it could lose the possibility of learning any hidden or unknown phases. On the contrary, the other one is called unsupervised learning and it requires no prior labeling and learns simply from the training data themselves. As a result, the unsupervised learning would be a more natural choice when one wants to explore the parameter space where known a priori is none or almost lack of.

In fact, tasks about identifying phases of matter by using unsupervised learning approach have been shown to be feasible as well as suggestive of new perspectives  \cite{wang16,Tanaka17,Hu17,Broecker17b,Wetzel17b,Chng18,Liu18,Durr19,Kottmann20}. For instance, algorithms such as autoencoders \cite{Lecun87,Kamp88,Hinton94} can extract out a local order parameter in the two dimensional (2D) Ising model \cite{Hu17,Wetzel17b}; clustering and dimensional reduction techniques such as diffusion maps have been employed to accurately distinguish from different topological sectors in the 2D XY model \cite{Scheurer19}. However, while these examples reflect the ``effectiveness'' of the unsupervised approach, the resulting phase boundaries are often not comparable to the supervised counterparts.

Therefore, we here propose to apply an unsupervised learning method for {\it identifying} (symmetry-protected) topological phases of matter, optionally followed by a ``supervised'' learning to further {\it determine} phase boundaries more accurately. We expand our previous work \cite{tsai20} on the one-dimensional $p$-wave superconductor (1D $p$-SC) \cite{kitaev01} and further on the Su-Schrieffer-Heeger (SSH) model \cite{Su79}, where both of the systems possess non-trivial topological phases. Moreover, unlike previous studies on similar systems \cite{Nieuwenburg17,Zhang18}, we emphasize the quantum information-based quantities as our input features for machine learning. In particular, we would focus on the block correlation matrices and Majorana correlation matrices, which have been approved effective according to Ref.~\onlinecite{tsai20}. We find that our proposed strategy not only reproduces similar results with sharp phase boundaries, but notably it does not rely on any prior knowledge of the phase space ({\it e.g.} number of phases present).

\section{Models} 
In this paper, we demonstrate the effectiveness of our proposed strategy on two classic models with topological phase transitions. The first one is Kitaev's 1D $p$-wave superconductor of spinless fermions \cite{kitaev01}:  
\bea  
H &= & \sum_{i}  -t \left (c_i^{\dagger} c_{i+1} + c_{i+1}^{\dagger} c_i \right)  \nonumber \\ 
&+& \Delta \left(c_i c_{i+1} + c_{i+1}^{\dagger} c_{i}^{\dagger} \right)  - \mu
\left( c_{i}^{\dagger} c_i -1/2\right),  
\label{H:pwave}
\eea
with the nearest-neighbor hopping amplitude $t$, superconducting  pairing potential $\Delta$, and on-site chemical potential $\mu$.   Due to translational invariance of $H$ it can be transformed into momentum space as
\be  \label{H:R} 
H = -\sum_{k \in BZ} \left (c_k^{\dagger}, c_{-k} \right ) 
\left[ {\mathbf R}(k) \cdot{\boldsymbol \sigma} \right] 
\left (c_k, c_{-k}^{\dagger} \right )^T, 
\ee 
where Pauli matrices ${\boldsymbol \sigma} = (\sigma_x, \sigma_y, \sigma_z)$, and ${\mathbf R}(k) =  (0, -\Delta \sin{k}, t\cos{k} + \mu/2)$ is the pseudo-magnetic field. One can compute its one-particle energy spectrum easily, which is given by $\epsilon (k) =  \pm \sqrt{(2t \cos{k} + \mu)^2 + 4\Delta^2 \sin^2{k}} $. 
The spinless $p$-wave superconductor (\ref{H:pwave}) breaks both time-reversal and chiral symmetries while keeps the particle-hole symmetry intact, and thus it belongs to the class D according to the ten-fold way classification for symmetry-protected topological systems \cite{Schnyder08}. The system ground state can be characterized by a $Z_2$ topological invariant. 

\begin{figure}[t]
	\begin{center}
		\includegraphics[width=7.5cm]{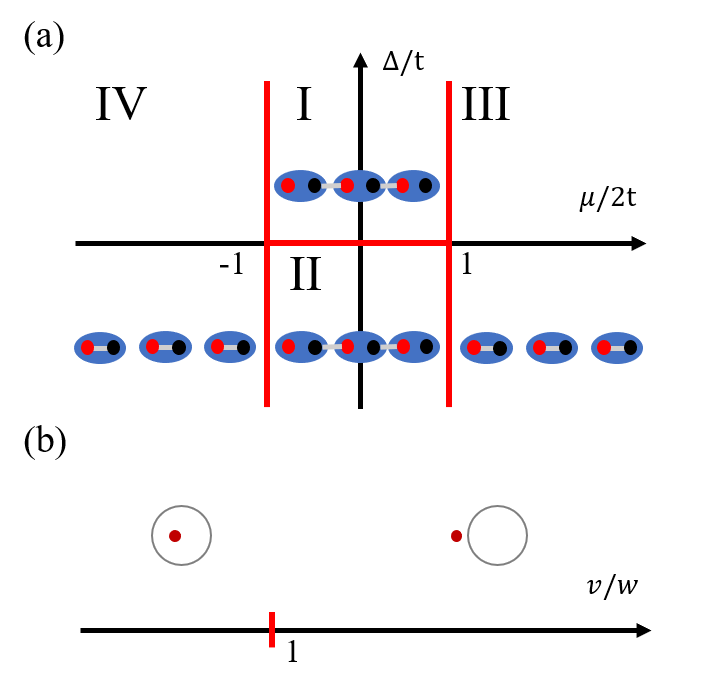}
		\caption{(a) Topological phase diagram of the 1D $p$-wave superconductor. The chain-like inset pictures schematically represent ground state for each phase in terms of Majorana fermions as described in the main context. The pairing between neighboring sites indicates phases I and II are topological, while the others are not. (b) Topological phase diagram of SSH model as a function of $v/w$ (lower part). Whether the trajectory of ${\mathbf R^\prime(k)}$ encloses the origin (red point) as $k$ runs over the first Brillouin zone determines if the corresponding phase is topological (upper part).}
		\label{fig:topological}
	\end{center}
\end{figure}

To make a clearer physical picture, we employ Majorana operators, $d_{2j-1} = c_j + c_j^{\dagger}$ 
and $d_{2j} = -i(c_j  - c_j^{\dagger})$, to rewrite Eq.~(\ref{H:pwave}) as 
\bea
H &=& \frac{i}{2} \sum_j [(-t+|\Delta|) d_{2j-1} d_{2j+2} + (t+|\Delta|) d_{2j}d_{2j+1}  \nonumber  \\
&-& \mu d_{2j-1} d_{2j}].
\label{H:majorana}
\eea
When $|\mu| > 2t$, Eq.~(\ref{H:majorana}) can be adiabatically transformed into the form, $\frac{-i\mu}{2}\sum_j d_{2j-1}d_{2j}$, where $t=|\Delta| =0$ and $\mu < 0$. As schematically depicted in Fig.~\ref{fig:topological}(a), the ground state of this simplified Hamiltonian is now composed of a paired Majorana fermions at the same site, leading to no Majorana edge modes and hence belongs to a topologically trivial phase (phases III and IV). In the opposite situation where $|\mu| < 2t$, Eq.~(\ref{H:majorana}) can be adiabatically transformed into a special case, $it \sum_j d_{2j} d_{2j+1}$, where  $t=|\Delta| >0$ and $\mu = 0$. The ground state in this case can then be viewed as follows: Most Majorana fermions from neighboring sites are paired together while the system leaves the edge Majorana modes alone (unpaired), and thus corresponding to a nontrivial phase [phases I and II in  Fig.~\ref{fig:topological}(a)]. 

The other model we have studied is the Su-Schrieffer-Heeger (SSH) model that describes spinless fermions on 1D lattice with two-site ($\alpha$ and $\beta$) unit cells at half-filling \cite{Su79}:
\be  
H = \sum_{i} v \left (c_{i, \alpha}^{\dagger} c_{i, \beta} + h.c. \right)  + w \left(c_{i+1, \alpha}^{\dagger} c_{i, \beta} + h.c. \right),
\label{H:SSH}
\ee
where $v$-terms represent fermion hopping within each unit cell $i$ and $w$-terms represent those hopping between nearest-neighbor unit cells. By transforming Eq.~(\ref{H:SSH}) into momentum space with periodic boundary condition, it becomes
\be  \label{H:SSH_R} 
H = \sum_{k \in BZ} \left (c_{k, \alpha}^{\dagger}, c_{k, \beta}^{\dagger} \right ) 
\left[ {\mathbf R^\prime}(k) \cdot{\boldsymbol \sigma} \right] 
\left (c_{k, \alpha}, c_{k, \beta} \right )^T, 
\ee 
where again ${\boldsymbol \sigma} $ denotes Pauli matrices and ${\mathbf R^\prime}(k) =  (v+w\cos k, w\sin k, 0)$. The eigenvalues and their corresponding eigenvectors can be easily represented by ${\mathbf R^\prime}(k)$. This model preserves all time-reversal, particle-hole, and chiral (sub-lattice) symmetries \cite{ssh_topological} 
and thus it belongs to the class BDI, characterized by $Z$ topological invariants in the symmetry-protected topological systems \cite{Schnyder08}. 

The topological nature of the SSH model can be intuitively understood by plotting the trajectory of ${\mathbf R^\prime}(k)$ over the first Brillouin zone. When $v/w < 1$ (topological), the trajectory winds about the origin (red point), while $v/w > 1$ (non-topological) it does not, as shown in the upper part of Fig.~\ref{fig:topological}(b). It turns out that this observation is closely related to the well-known winding number \cite{Sheet20} and results in a phase diagram schematically shown in the lower part of Fig.~\ref{fig:topological}(b).

\begin{figure}[t]
	\begin{center}
		\includegraphics[width=8.0cm]{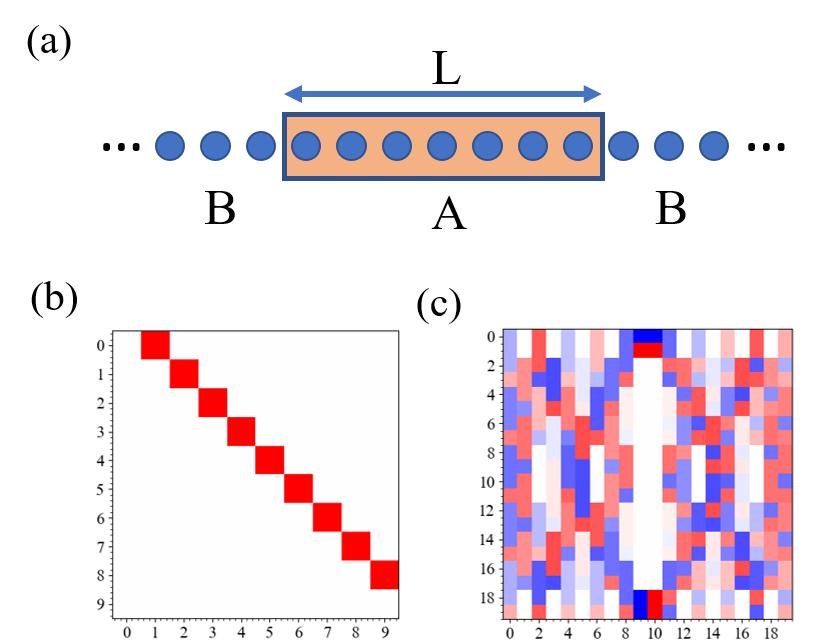}
		\caption{(a) The infinite system is composed of a finite subsystem A with $L$ sites and an environment B. (b) A typical $L\times L$ MCM image for a 1D $p$-wave SC at $\Delta/t=1,\mu=0, L=10$. (c) A typical $2L\times 2L$ eigenvector image of a BCM for the same system.}
		\label{fig:subsystem}
	\end{center}
\end{figure}

\section{Methods of Machine Learning}

\subsection{Producing data}
As have been mentioned in the Introduction section, ML is a data-driven approach. Any effective ML pipeline should be qualified by training data, a model architecture, and a fair evaluation procedure. Due to the essential role of data, here, we follow our previous successful experience on using entanglement correlations as representative data for our interested systems \cite{tsai20}.

We consider two common entanglement-based correlators. Both of them calculate certain correlations within a subsystem A after integrating out all the other degrees of freedom outside A, {\it i.e.}, its environment B [see Fig.~\ref{fig:subsystem}(a)]. First, the ``Majorana'' correlation matrix (MCM) for fermions at two sites within the subsystem A of size $L$ can be defined, more concretely, via Majorana language for a 1D $p$-wave SC ($t\equiv 1$), 
\bea
\label{eq:mcm}
& i& \tr \rho_0 d_{i} d_{j} = \tr \rho_0 (c_i - c_i^{\dagger})(c_j +c_j^{\dagger}) = \\
&&\int_{0}^{\pi}\frac{dk}{\pi}\frac{-\Delta \sin k \sin[k(i-j)]+(\cos k +\frac{\mu}{2})\cos [k(i-j)]}{\sqrt{\Delta^2\sin^2 k + (\cos k +\frac{\mu}{2})^2}}, \nonumber
\eea
where $\rho_0$ represents the density matrix of the ground state and $i, j$ represent two sites within A. A typical MCM ``image'' is shown in Fig.~\ref{fig:subsystem}(b). When turning into an insulating case such as the SSH model, MCM would have to be modified as follows ($w \equiv 1$),
\bea
\label{eq:scm}
&& \tr \rho_0 c_{i, \beta} c^{\dagger}_{j, \alpha} = \\
&&\int_{0}^{2\pi}\frac{dk}{\pi}\frac{\sin k \sin[k(i-j)]+(v + \cos k)\cos [k(i-j)]}{\sqrt{1+ v^2+ 2v\cos k}}, \nonumber
\eea
where $i, j$ indicate two unit cells within A. Notably, the particle-hole space should now be replaced by sub-lattice space, but for simplicity we still call it MCM. 

The second type of the correlators is block correlation matrix (BCM) for subsystem A, defined as 
$\text{BCM}_{i,j} = \tr \rho_0 {\hat{\mathbf{c}}}_i \hat{{\mathbf{c}}}_j^{\dagger} $ 
with $\hat{{\mathbf{c}}}_i \equiv (c_i, c_i^{\dagger})^T$ [ $(c_{i, \alpha}, c_{i, \beta})^T$] for 1D $p$-wave SC case (SSH model), and $i,j$ being sites (or unit-cells) of the finite subsystem A. This matrix is intimately connected to the more familiar quantity, the reduced density matrix of the block A,
$\rho_A = \bigotimes_m
\left[\begin{matrix} \lambda_m & 0\\ 0 & 1-\lambda_m \end{matrix}
\right]$, 
where $\lambda_m$ are simply the eigenvalues of BCM and $\lambda_m$s are also known as one-particle entanglement spectrum (OPES). Therefore, the eigenvalues and their corresponding eigenvectors of BCM, also known as one-particle entanglement eigenvectors (OPEEs), would be considered as our input data (``image'') for our ML purpose.  Fig.~\ref{fig:subsystem}(c) provides an example of the eigenvector ``image'' from a given BCM.

\begin{figure}[t]
	\begin{center}
		\includegraphics[width=8.5cm]{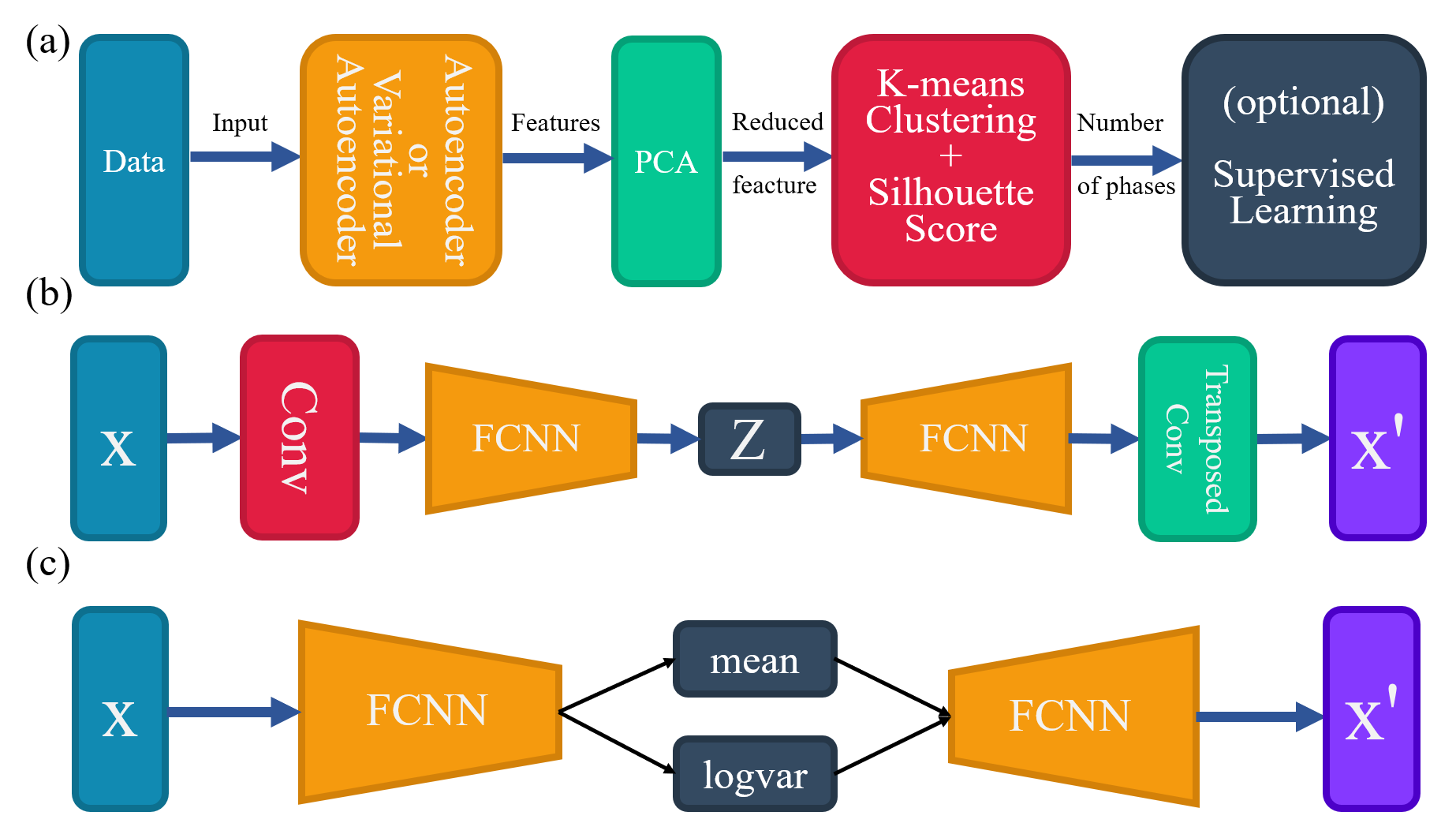}
		\caption{(a) The proposed working pipeline to identify different phases and to finely determine the phase boundaries without prior knowledges. The model architectures are schematically shown in (b) the autoencoder (AE) and (c) the variational autoencoder (VAE). [Conv: convolutional module; FCNN: fully connected neural network module; mean: mean values; logvar: logarithmic of the variances]}
		\label{fig:process}
	\end{center}
\end{figure}

\subsection{ML algorithms}
Once the format of our input data was settled down, we can then build our ML pipeline for our task, to identify topological phase transitions of a given system in an unsupervised manner without prior knowledges. Our proposed learning procedure integrates a few different ML algorithms into getting final predictions. There are four steps: (1) As shown in Fig.~\ref{fig:process}(a), the input data are first fed to an autoencoder for training \cite{Lecun87,Kamp88,Hinton94} in order to extract effective features; (2) the number of necessary features is then determined by principle component analysis (PCA) \cite{Pearson01,Jolliffe02} once $99\%$ (a prescription) of the total variance of input features is kept \cite{prescription}; (3) it turns out that the total number of phases in the focused parameter space can be determined by K-means clustering \cite{MacQueen67,Lloyd82} from the transformed features after PCA, followed by Silhouette analysis (SA) \cite{Rousseeuw87,deAmorim15}; (4) finally, relatively sharp phase boundaries can be determined with the help of supervised learning: Constructing a training dataset by expanding around the data point with the highest confidence in each cluster using the same SA, a neural network can then be trained to achieve the goal. We emphasize that step (4) is optional and not always necessary for our purpose of use. In the following, we provide a brief review for each algorithm, and refer readers to references \cite{GBC,Mehta19} for more details.

\begin{figure*}[t]
	\begin{center}
		\includegraphics[width=14.0cm]{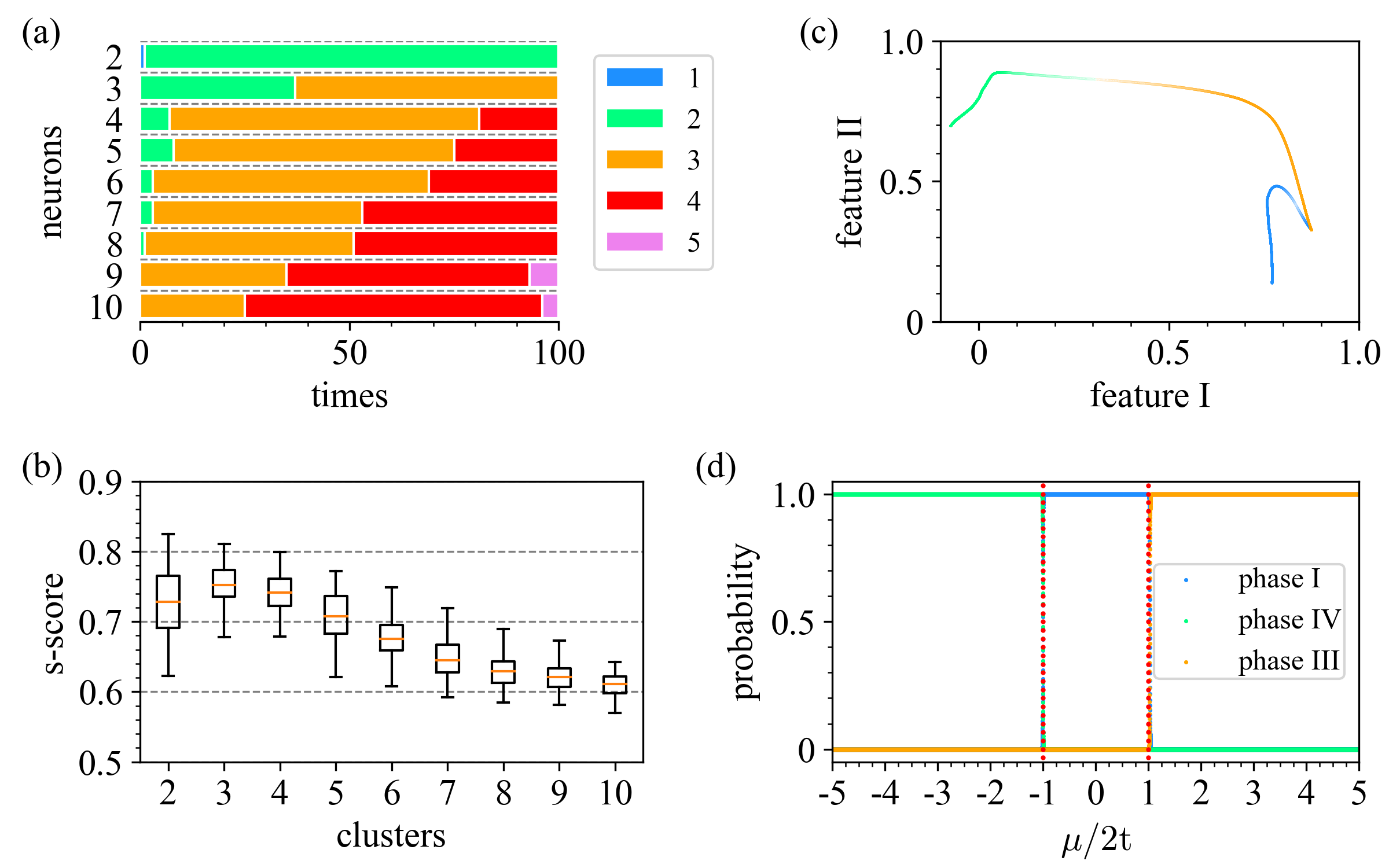}
		\caption{AE results for type-I data (MCMs). (a) The discrete distribution of necessary number of neurons $d_z$ for a given $n_{mid}$ neurons in the middlemost layer (2 to 10 along $y$-axis). Results from 100 independently trained AEs with type-I data are statistically calculated: The length of every color bar is proportional to the number of times that $d_z$ occurred within 100 models. Different color in the legend represents different $d_z$. (b) The box plot of the $s$-score as a function of $n$-clustering (via K-means method). (c) Latent representations projected to a subspace spanned by the first two principle components. Each color indicates its corresponding cluster (phase). (d) The neuron output ``phase diagram'' as a function of $\mu/2t$ with $\Delta/t=1, L=10$ for 1D $p$-SC from a trained CNN by supervised learning in the last step of the ML pipeline.}
		\label{fig:AE_1DSC_BA}
	\end{center}
\end{figure*}

\subsubsection{Autoencoder and its variational version}
An autoencoder (AE) is a type of neural network that aims at compressing input data into more efficient representation in an unsupervised manner. It mainly consists of two parts, an encoder and a decoder. The encoder $f$ takes $d$-dimensional input data $x$ and outputs a $l$-dimensional latent variable $z = f(x)$; the decoder $g$ then maps $z$ back to $x^\prime = g(z)$ in $d$-dimension. The learnable parameters of the model are trained by performing gradient descent updates in order to minimize the reconstruction loss, $L(x, x^\prime)$, usually chosen to be the mean squared error. Since  typically $l<d$ after encoding the data, using AE is often viewed as a {\it nonlinear} dimension reduction method. 

A typical model architecture we used in this study is schematically shown in Fig.~\ref{fig:process}(b). In the encoder part, it is made of a convolutional module (ResNet-like structure \cite{He16}) followed by a linear module consisting of a few fully connected hidden layers. In the decoder part, it is basically made of similar layer structures but arranged in a reversed order with respect to the encoder. Note that, however, the convolutional layers here are replaced by transposed convolutional ones. The activation functions of the intermediate layers are always rectified linear units (ReLUs)  \cite{Jarrett09}, except for the final layer where the sigmoid function is used.

Although the traditional AE can learn a function to encode two input data points into distinct latent variables $z_1$ and $z_2$, one may have no idea what would be the decoded result when giving the input $(z_1+z_2)/2$. To overcome this arbitrariness in AEs, we also consider the variational autoencoders (VAE) which can learn a latent variable model $g(x, z)$ with a joint distribution of a latent variable $z$ and input $x$ \cite{Kingma13}. In sharp contrast with traditional AEs, $z$ here is basically drawn from some prior probability distribution $p(z)$, which is almost always chosen to be a multivariate Gaussian, and thus leads to certain controllability. In addition, the weights of the VAE are now trained by simultaneously optimizing two loss functions, a reconstruction loss and the Kullback-Leibler (KL) divergence between the learned latent distribution and a prior unit Gaussian. Such an additional KL divergence loss can be viewed as a regularization term in a traditional AE. The VAE has a similar model architecture compared to that in Fig.~\ref{fig:process}(b) by simply getting rid of convolutional modules at the head and tail. Furthermore, as shown in Fig.~\ref{fig:process}(c) the middle most layer would output the multiple means and the variances, depending on the number for encoded features we need, in the latent space.

\subsubsection{Principle component analysis}
Principal component analysis (PCA) is a standard yet simple method for dimensional reduction and data visualization \cite{Pearson01,Jolliffe02}. It is an {\it orthogonal}, {\it linear} transformation of the input features to a sorted set of new variables by their variance. Such method is motivated by the experience that in many cases, the most relevant information can be revealed in the direction with largest variance for a given signal, while directions with small variance usually indicate noises and may be neglected.

Concretely, let us consider $N$ $p$-dimensional feature vectors, ${\bs X} = \{{\bs x}_1, {\bs x}_2,\cdots,{\bs x}_N\}$. One can assume that the mean of all vectors, $\sum_i{\bs x}_i = 0$, without loss of generality, and then ${\bs X}$ is called a (zero-mean) centered matrix. By definition, the transformation weight vector for producing the first principle component ${\bs w}_1$ can be found by
 \be
 {\bs w}_1 = \text{argmax}_{||w||=1} \lbrace \sum_i({\bs x}_i\cdot{\bs w})^2\rbrace .
  \label{eq:pca}
 \ee
The next ordered weight vectors are then obtained by repeating Eq.~(\ref{eq:pca}) after subtracting out the calculated principle components from ${\bs X}$. However, in practice, one can prove that this procedure for getting ${\bs w}_i$ is equivalent to find out the eigenvectors of the $N\times N$ symmetric matrix ${\bs X}^T{\bs X}$, \cite{Mehta19}{\it i.e.},
 \be
{\bs X}^T{\bs X}{\bs w}_i = \lambda_i{\bs w}_i,
\label{eq:pca_w}
\ee
where we have assumed that the obtained eigenvalues are sorted such that $\lambda_1\geq\lambda_2\geq\cdots\geq\lambda_N\geq 0$, representing variances for the input feature vectors. It is also useful to define the relative variance, $\tilde{\lambda}_i = \lambda_i/\sum_i{ \lambda_i}$ in order to count accumulated variance percentage.

\subsubsection{K-means clustering}
K-means clustering is a simple and easily understandable clustering algorithm without any supervision \cite{MacQueen67,Lloyd82}. Given a prior knowledge about the number of clusters $K$, the basic idea is to find the best cluster means such that the variance within each cluster is minimized. To put it more precisely, consider a set of $N$ $p$-dimensional data points without labels, ${\bs X} = \{{\bs x}_1, {\bs x}_2,\cdots,{\bs x}_N\}$ and call the set, ${\bs C} = \{{\bs \mu}_1, {\bs \mu}_2,\cdots,{\bs \mu}_K\}$ (${\bs \mu}$ is also $p$-dimensional), as the $K$ centers for the whole data. The objective of K-means method is then to assign each ${\bs x}_i$ to an appropriate cluster such that the loss function, 
 \be
L(\{{\bs X}, {\bs C}\}) = \sum_{k=1}^{K}\sum_{i=1}^{N} a_{ik}({\bs x}_i-{\bs \mu}_k)^2,
\ee
is minimized. Note that the assignment $a_{ik}$ is 1 if ${\bs x}_i$ is assigned to cluster $k$ while 0 otherwise and $\sum_k a_{ik} = 1$ for every $i$. The implementation is usually done by iteration until certain convergence with chosen tolerance level has been achieved.

\subsubsection{Silhouette analysis}
Although K-means clustering is quite intuitive, one still needs to provide the number of clusters $n$ as a priori knowledge. To obtain a more reasonable estimation of this number and somehow eliminate the effect of distance function chosen in K-means method, we employ Silhouette analysis (SA) to justify it \cite{Rousseeuw87,deAmorim15}. For a given set of clusters $\{\mathcal{C}_i\}$, SA assigns a value to each data point $x\in \mathcal{C}_i$ by
\be
s(x) = \frac{b(x)-a(x)}{\text{max}\{b(x), a(x)\}},
\ee
where $b(x) = \min_{j\neq i} b_j(x)$ with $b_j(x)$ being the mean distance of $x$ to all points in the cluster $j$ and $a(x)$ is the mean distance between $x$ and all other data points in the same cluster $i$. In other words, the Silhouette value $s(x)$, bounded between $\pm 1$, is a measure of how similar $x$ is to its own cluster (cohesion) compared to the other clusters (separation). Considering the mean of Silhouette values, $s$-score, as a function of $n$ (after K-means clustering), the best estimation of $n$ is simply the one gives maximum of $s$-score.

In addition, once the best $n$ is determined, the data point with the highest Silhouette value within each cluster may then be taken as a confident seed to build a ``labeled'' training set to train a neural network via supervised learning approach. It in turn could be used to make sharper phase boundaries in a phase diagram, originally obtained in an unsupervised way over interested parameter space.

\begin{figure}[tb]
	\begin{center}
		\includegraphics[width=8.0cm]{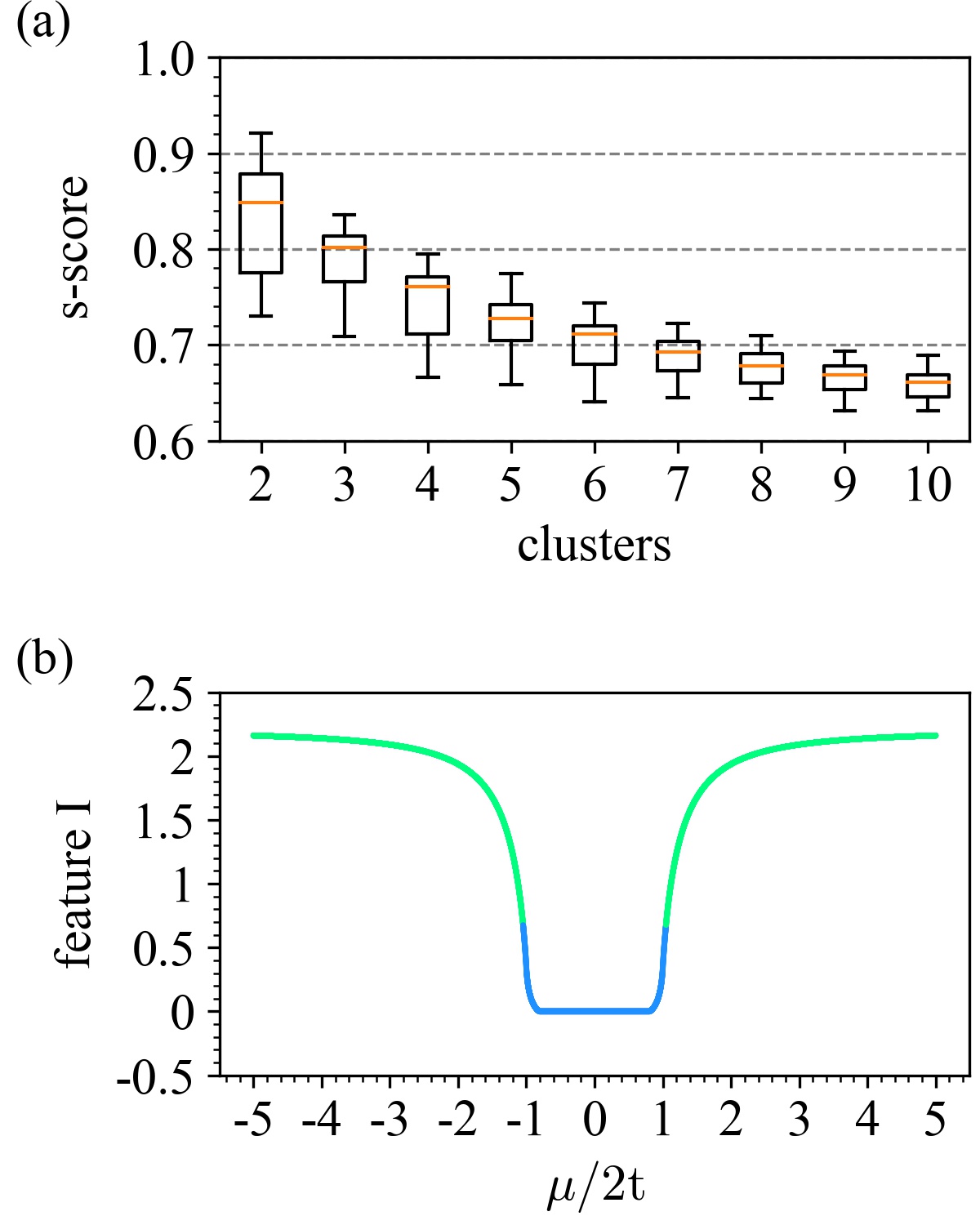}
		\caption{AE results for type-II data (OPES). (a) The box plot of the $s$-score as a function of $n$-clustering (via K-means method). (b) Latent representations (of type-II data) after projected along the first principle component of PCA, as a function of $\mu/2t$. Each color indicates its corresponding cluster (phase).}
		\label{fig:AE_1DSC_BCMev}
	\end{center}
\end{figure}

\begin{figure*}[tb]
	\begin{center}
		\includegraphics[width=15.0cm]{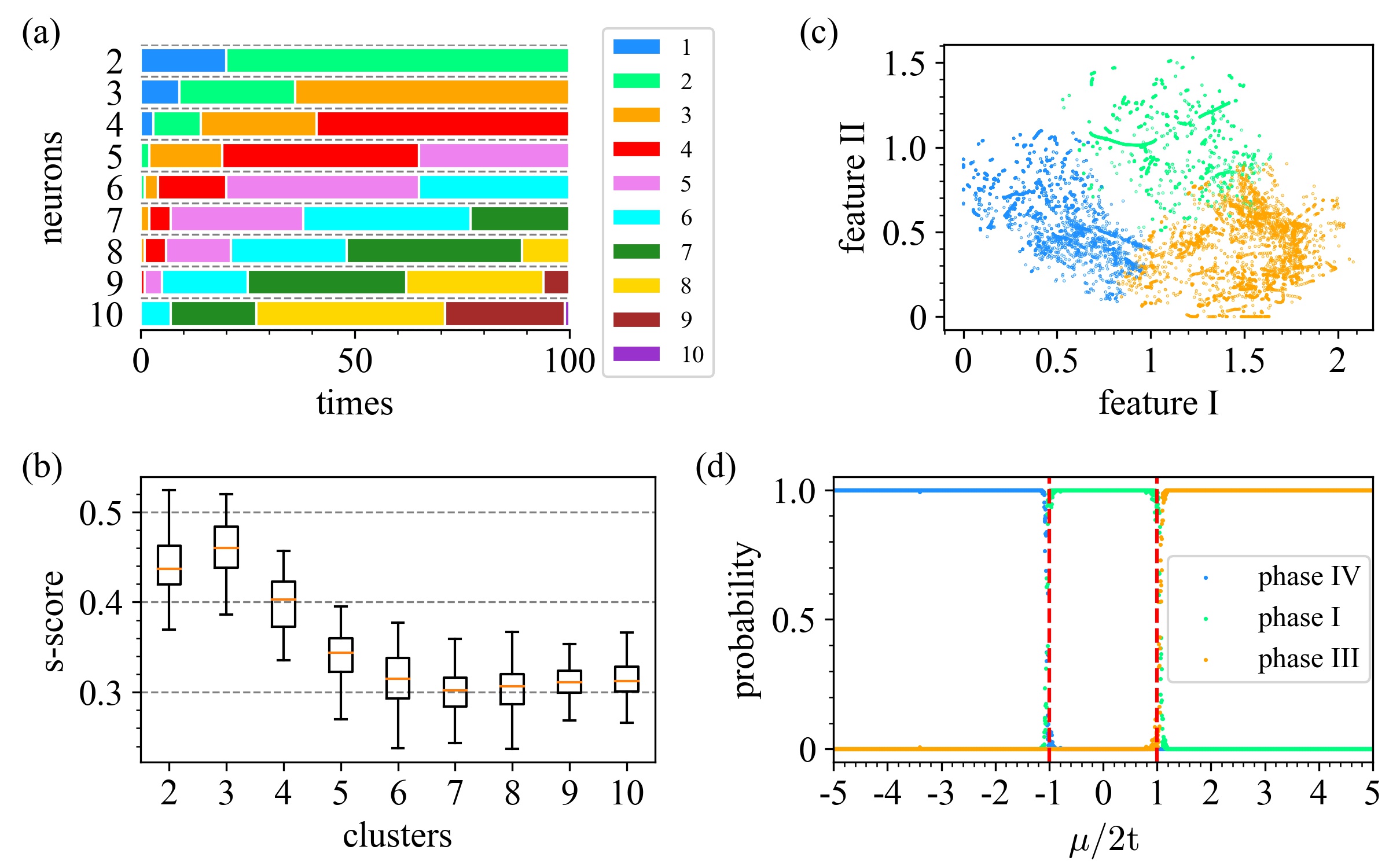}
		\caption{AE results for type-III data (OPEEs). (a) The discrete distribution of necessary number of neurons $d_z$ for a given $n_{mid}$ neurons in the middlemost layer (2 to 10 along $y$-axis). Results from 100 independently trained AEs with type-III data are statistically calculated: The length of every color bar is proportional to the number of times that $d_z$ occurred within 100 models. Different color in the legend represents different $d_z$ (1 to 10 here). (b) The box plot of the $s$-score as a function of $n$-clustering (via K-means method). (c) Latent representations after projected to a subspace spanned by the first two principle components. Each color indicates its corresponding cluster (phase). (d) The neuron output ``phase diagram'' as a function of $\mu/2t$  with $\Delta/t=1, L=100$ from a trained CNN by supervised learning.}
		\label{fig:AE_1DSC_BCMevec}
	\end{center}
\end{figure*}

\section{Results and Analysis}
\subsection{1D $p$-wave superconductor}
We first prepare the input “image” dataset by generating 20,001 MCMs via Eq.~(\ref{eq:mcm}) at evenly divided $\mu/2t$ from -5 to 5, with subsystem size $L$ (block A) of an infinite chain with periodic boundary conditions. Each MCM can be viewed as a $L \times L$ “image” in one (gray) channel and entries in it represent pixel values. We call this type-I input format. Without loss of generality, we will assume $\Delta/t=1$, $L=10$, and $2t\equiv 1$ (energy units) unless mentioned otherwise.

The other formats of the input dataset could originate from BCM, as mentioned in Sec. III A. For a finite subsystem A of size $L$, we prepare again 20,001 BCMs (now of size $2L \times 2L$ due to Nambu notation) at evenly divided $\mu/2t$ from -5 to 5. In our study, we either collect all eigenvalues of each BCM as an input vector (called type-II input) or arrange each eigenvector of a BCM as one of the columns in a new matrix $M$ (of size $2L \times 2L$), viewed as a ``gray image'' (called type-III input).

\subsubsection{AE approach}
Following our ML pipeline mentioned in Section III, the first step is to train a neural network to encode our type-I input data to effective representations in the latent space.  However, in order to determine the minimal dimension $d_z$ of the latent space, we train a series of AEs with same model architecture \cite{appendix} except for the number of hidden neurons in the middlemost layer  ($n_{mid}$, from 2 to 10).  For each $n_{mid}$, we record the necessary dimension of the (converged) latent representations to keep at least $99\%$ variance of them by PCA. Due to the unconstrained nature of the latent representations in AEs, we repeat 100 times of the same training procedure with the same initial weight distribution. As shown in Fig.~\ref{fig:AE_1DSC_BA}(a), we observe that the minimal dimension $d_z$ would be 4, because such number becomes dominant in the discrete distribution of $d_z$ when $n_{mid}$ increases.

Next, SA is utilized to estimate the best number of clusters $n$ for the latent representations of all input data via K-means method. Note that these representations are provided from previous trained AEs with $n_{mid} = 4$ (as suggested above) hidden neurons in the middlemost layer. The box plot,  Fig.~\ref{fig:AE_1DSC_BA}(b), clearly shows that the mean of Silhouette value achieves the highest one when $n=3$. By projecting the 4D latent representations into a 2D space spanned by the first two principle components (features), we obtain Fig.~\ref{fig:AE_1DSC_BA}(c). This feature plot gives us an insight about how the system could be reasonably divided into three clusters (phases).

\begin{figure}[tb]
	\begin{center}
		\includegraphics[width=7.5cm]{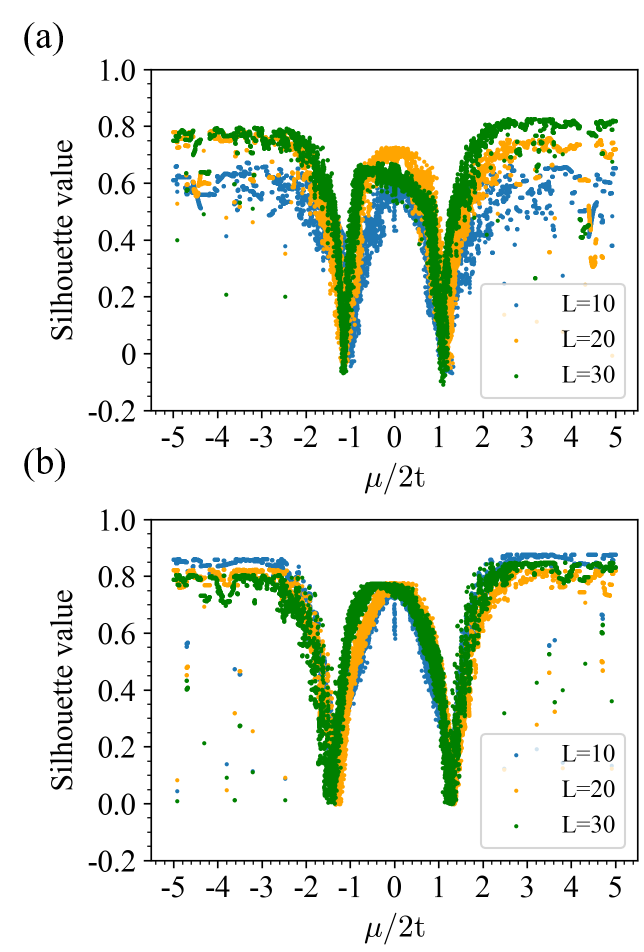}
		\caption{ Silhouette values as a function of $\mu/2t$ with various subsystem size $L$ using (a) the AE approach and (b) the VAE approach, given the type-III (OPEEs) training data after feature extraction. It is remarkable to see that all the dips reasonably indicate the phase (cluster) boundaries. Moreover, the highest value corresponding to each cluster only change mildly as $L$ increases.}
		\label{fig:AE_1DSC_BCMevec_score}
	\end{center}
\end{figure}

Again, due to the unconstrained nature of the latent representations in AEs, the phase transition points found are statistically at mean values -1.012 and 0.980 with standard deviations 0.093 and 0.081, respectively, after collecting clustering results from 100 sets of latent representations via different trained AEs. This is quite close to the theoretically calculated values -1 and 1, but having relatively large deviation. 
To make phase boundaries sharper, we further train a CNN classifier \cite{appendix} in supervised learning manner to predict the whole phase diagram. To prepare a ``labeled'' dataset, we first pick up three seeds in the input ``images'', each of which gets the highest Silhouette value in each cluster and is thus believed to be inside each phase with strong confidence. In our example, the three seeds are located at $\mu/2t=-2.988, 0.182, 3.019$ (from latent representations traced back to original input data points), and for each cluster (phase) we expand symmetrically around the seed by a window width $0.1t$ to obtain 2000 points with equal spacing. These 6000 data points are finally formed our training dataset, while the original 20,001 ones become our test set without labels. As shown in Fig.~\ref{fig:AE_1DSC_BA}(d), the phase boundaries obtained by trained CNN classifiers are clearly sharper at mean values -1.015 and 1.038 with smaller deviations 0.018 and 0.023 (after repeating same training procedure 100 times). 

\begin{figure}[tb]
	\begin{center}
		\includegraphics[width=7.5cm]{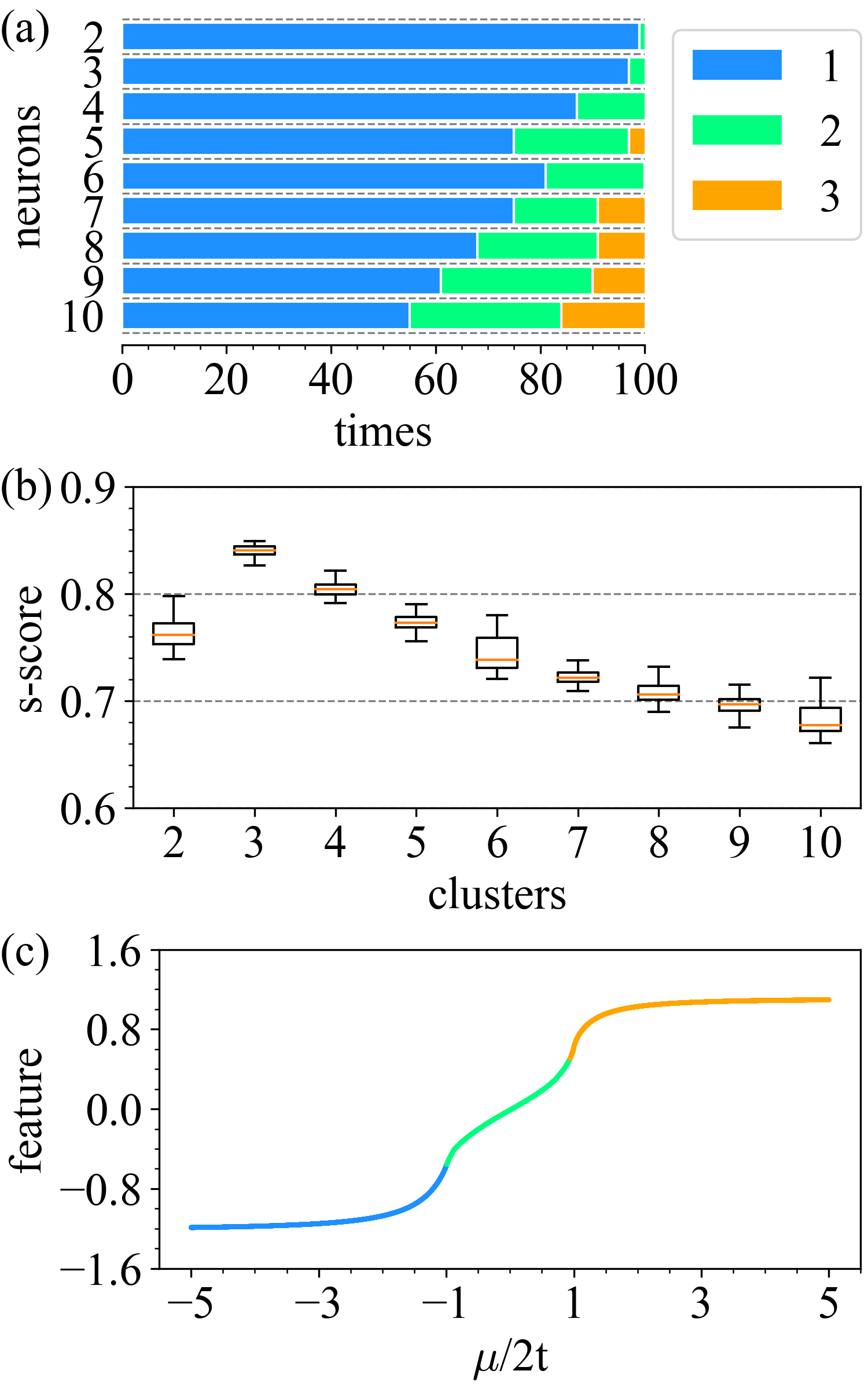}
		\caption{VAE results for type-I data (MCMs). (a) The discrete distribution of necessary number of neurons $d_z$ for a given $n_{mid}$ (paired) neurons in the middlemost layer (2 to 10 along $y$-axis). Results from 100 independently trained VAEs are statistically calculated: The dominant $d_z$ is clearly one (pair) here. (b) The box plot of the $s$-score as a function of $n$-clustering (via K-means method). (c) Latent representations (of type-I data) as a function of $\mu/2t$. Each color indicates its corresponding cluster (phase).}
		\label{fig:VAE_1DSC_BA}
	\end{center}
\end{figure}

Alternatively, we next consider taking BCM generated quantities as our training inputs. There are two potential formats. We first train an AE to encode the type-II inputs, whose format looks simpler. Note that here we choose the middlemost layer to have 2 hidden neurons according to similar experiments done in type-I case [(see Fig.~\ref{fig:AE_1DSC_BA}(a)]; moreover, all convolution-related modules are taken away in the model architecture \cite{appendix}. 
As shown in Fig.~\ref{fig:AE_1DSC_BCMev}(a), SA shows that 2 clusters, among other chosen number of distinct clusters, are the best result obtained via K-means method on the latent representations. It indicates that, when combining with the projected plot of the latent features, as seen in Fig.~\ref{fig:AE_1DSC_BCMev}(b), this approach can only distinguish a topological phase (phase I) from non-topological ones (phases III and IV). The reason that phases III and IV can not be further distinguished can be attributed to rather limited information provided by type-II inputs, as indicated in our previous work \cite{tsai20}.

To gain more information, we use type-III inputs to train autoencoders with $n_{mid}$ hidden neurons in the middlemost layer to get the corresponding latent representations. As shown in Fig.~\ref{fig:AE_1DSC_BCMevec}(a), the necessary dimension of the latent space is proportional to $n_{mid}$ and thus we choose the dominant one at large $n_{mid}$, namely, $d_z=8$. Note that to avoid the arbitrariness of phase when computing eigenvectors, we have preprocessed the inputs by squaring each entry of the input matrix (``image''). After K-means clustering the latent representations obtained from AE, Fig.~\ref{fig:AE_1DSC_BCMevec}(b) depicts the results of SA and the input data are suggested to be separated into 3 clusters. Fig.~\ref{fig:AE_1DSC_BCMevec}(c) shows how the encoded representations in the 3D latent space after projecting to 2D may be divided into 3 clusters. In fact, by plotting the phase diagram as a function of $\mu$, the phase transition boundaries are somewhat shifted from theoretical values with relatively large deviation. This, however, can be improved if following the same supervised learning strategy as mentioned in the type-I input case \cite{typeiii}. Note that the labeled training dataset used here is based on 3 seeds at $\mu/2t=-3.362, -0.026, 3.359$, each of which gets the highest Silhouette value in the corresponding cluster.  Even though these vales are taken in the subsystem A with $L=10$, Fig.~\ref{fig:AE_1DSC_BCMevec_score}  shows that the locations of the most confident Silhouette values do not change much as $L$ increases. Therefore, we finally take the subsystem of size $L=100$ to refine the transition boundaries and to reduce the possible finite-size effect all together. As shown in Fig.~\ref{fig:AE_1DSC_BCMevec}(d), the phase transition points found then are statistically at mean values -0.917 and 0.978 with standard deviations 0.149 and 0.0719, respectively, using well-trained CNN classifiers \cite{appendix}.

\begin{figure}[tb]
	\begin{center}
		\includegraphics[width=7.5cm]{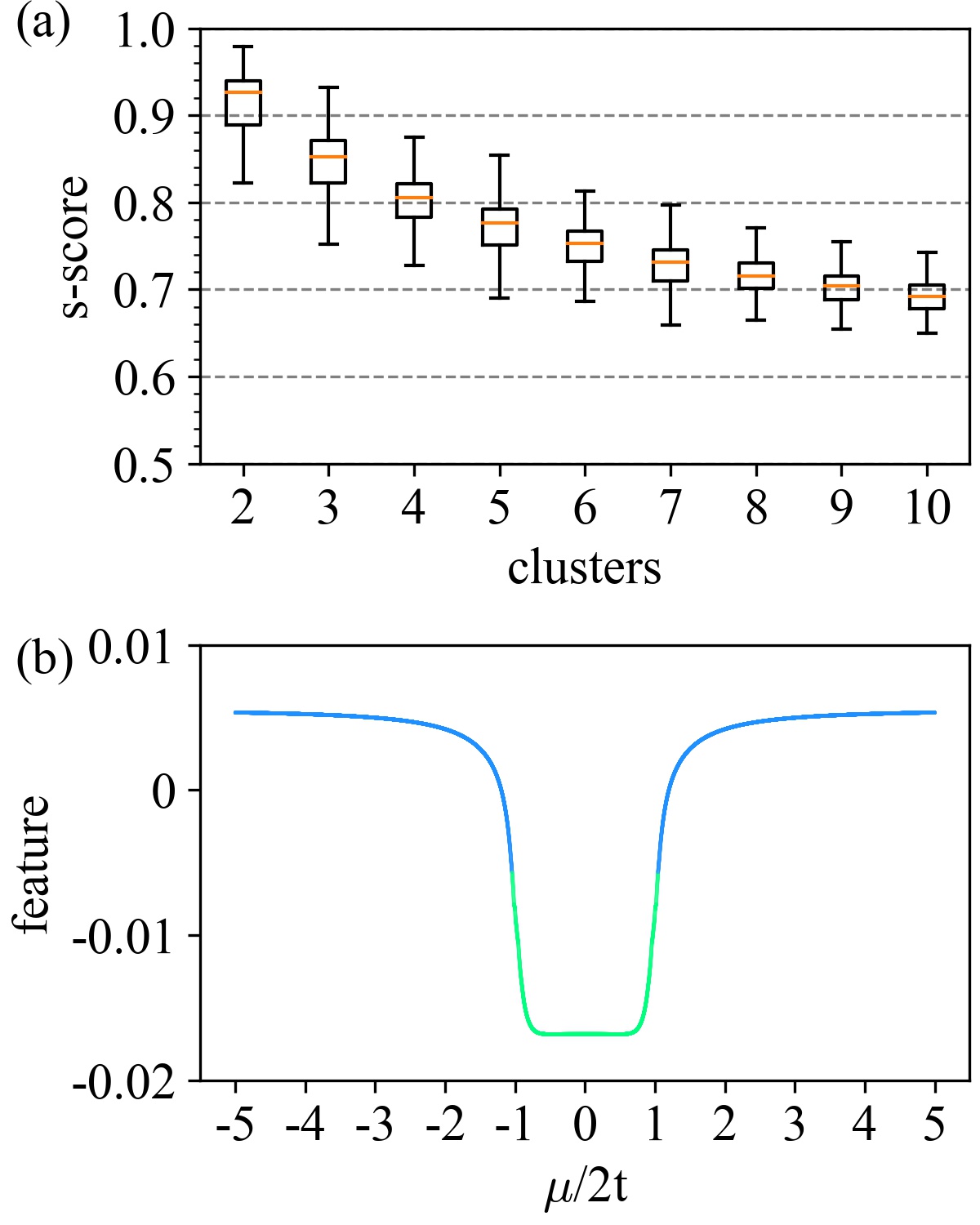}
		\caption{VAE results for type-II data (OPES). (a) The box plot of the $s$-score as a function of $n$-clustering (via K-means method). (b) Latent representations (of type-II data) extracted by a trained VAE as a function of $\mu/2t$. Each color indicates its corresponding cluster (phase).}
		\label{fig:VAE_1DSC_BCMev}
	\end{center}
\end{figure}

\subsubsection{VAE approach}
In VAE approach, we try to impose certain constraints (multivariate Gaussian here) on the encoded representations and let a VAE learn the parameters of a probability distribution modeling the input data. Following the same ML pipeline, the first step is to train a series of VAEs with $n_{mid}=2$ to 10 pairs of hidden neurons in the middlemost layer \cite{appendix}; in each pair one neuron outputs the mean of encoded representation distribution and the other produces its corresponding variance. After repeating the same training procedure 100 times for a given $n_{mid}$, the number of necessary dimension for the latent representations obtained after PCA to keep at least $99\%$ variance of the encoded data is simply one (pair). In sharp contrast with AE, VAE provides a more stable result, as clearly shown in Fig.~\ref{fig:VAE_1DSC_BA}(a).

Once the minimal dimension of the encoded representations is set, we employs SA to estimate the best number of clusters $n$ for them via K-means method. From Fig.~\ref{fig:VAE_1DSC_BA}(b), it shows that the Silhouette value achieves the highest one when $n=3$. Fig.~\ref{fig:VAE_1DSC_BA}(c) depicts the encoded representations as a function of $\mu/2t$ and different colors indicate distinct clusters (phases). By plotting the phase diagram, we find that transition points are at mean values -0.992 and 0.9597 with standard deviations 0.047 and 0.026, respectively, after collecting results from 100 trained models with the same initial weight distribution. This small deviation is significantly different from AE approach, where the position of transition points found is more unstable. The supervised learning in the last step is not necessary in this case.

On the other hand, we take BCMs as our training input. Firstly, we train a VAE to encode the type-II inputs. Note that here we enforce the middlemost layer to output one set of mean and variance because similar PCA tests have been done to determine the minimal dimension of the encoding representations. Next, as shown in Fig.~\ref{fig:VAE_1DSC_BCMev}(a), SA shows that 2 clusters are the best result obtained through K-means method on the encoded representations [see Fig.~\ref{fig:VAE_1DSC_BCMev}(b)]. It turns out that this approach can distinguish topological phase (phase I) from non-topological ones (phases III and IV), but phases III and IV can not be further distinguished. This result is again consistent with previous work \cite{tsai20}, indicating too compressed information provided by type-II inputs.

\begin{figure}[tb]
	\begin{center}
		\includegraphics[width=7.5cm]{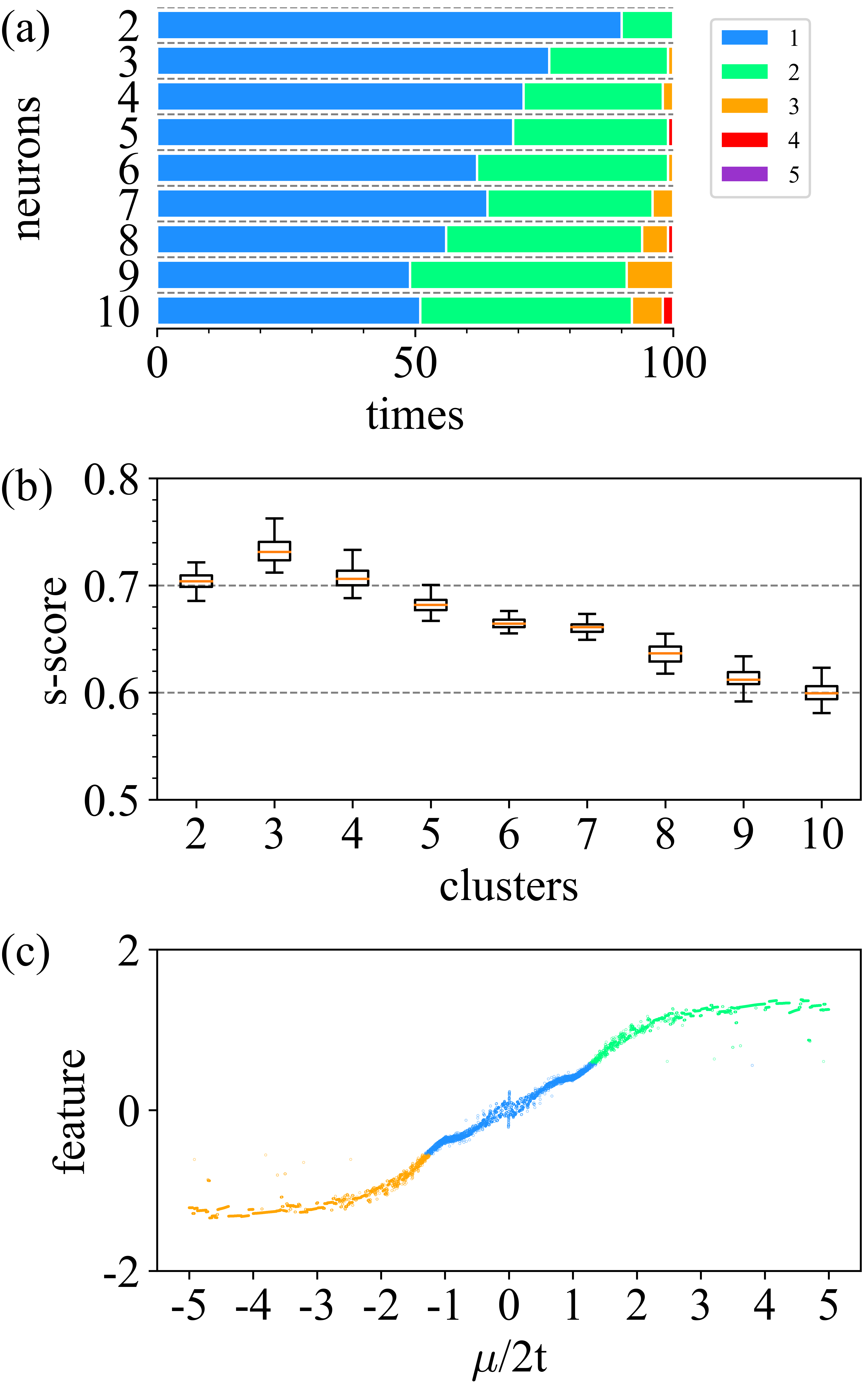}
		\caption{VAE for type-III data (OPEEs). (a) The discrete distribution of necessary number of neurons $d_z$ for a given $n_{mid}$ (paired) neurons in the middlemost layer (2 to 10 along $y$-axis). Results from 100 independently trained VAEs are statistically calculated: The dominant $d_z$ is clearly one (pair) here. (b) The box plot of the $s$-score as a function of $n$-clustering (via K-means method). (c) Latent representations (of type-III data) as a function of $\mu/2t$. Each color indicates its corresponding cluster (phase).}
		\label{fig:VAE_1DSC_BCMevec}
	\end{center}
\end{figure}

Thus, we next examine to train another VAEs by feeding in type-III inputs, which have been preprocessed by squaring each entry in matrices. We find the minimal dimension of the latent dimension to be 1 based on PCA test, as shown in Fig.~\ref{fig:VAE_1DSC_BCMevec}(a). After K-means clustering the latent representations, Fig.~\ref{fig:VAE_1DSC_BCMevec}(b) depicts the results of SA and the input data are suggested to be separated into 3 clusters. Fig.~\ref{fig:VAE_1DSC_BCMevec}(c) shows how the encoded representations in the latent space can be divided into 3 clusters. In fact, by drawing the output probability for each phase as a function of $\mu$, the phase transition points are somewhat deviated from theoretical values.
This, however, can be largely improved if following the same supervised learning strategy as in the MCM case \cite{typeiii} .  This way not only shifts transition points back to mean values -0.995 and 1.003, but also reduces standard deviation from 0.335, 0.244 to 0.107, 0.067, respectively (statistically over 100 same-architecture CNN classifiers). The ``labeled'' training dataset used here is based on 3 seeds at $\mu/2t=-3.375, -0.063, 2.892$, each of which gets the highest Silhouette value in the corresponding cluster. Note that, similar to the AE case, we again take the subsystem of size $L=100$ to refine the transition boundaries and to reduce the possible finite-size effect as well. 

\subsection{SSH model}
We prepare the input “image” dataset by generating 10,001 MCMs according to Eq.~(\ref{eq:scm}) at evenly divided $v/w$ from 0 to 10, with subsystem size $L$ (block A) under periodic boundary conditions of the full system. Note that the ground state of SSH model is basically an insulating state, therefore, ``MCM'' is called for convenience and is nothing to do with ``Majorana''. Each MCM can be viewed as a $L \times L$  “gray image” and entries in it represent pixel values. This forms our type-I input format and we will assume $w\equiv 1$ (energy units) and $L=10$. Furthermore, one can also prepare eigenvalues and eigenvectors of BCMs (now of size $2L \times 2L$ due to sublattice space), corresponding to type-II and type-III input formats, respectively. However, they do not bring new physics other than that from type-I format in unsupervised learning, and thus we omit the results for simplicity.

\subsubsection{AE approach}
\begin{figure*}[tb]
	\begin{center}
		\includegraphics[width=15.0cm]{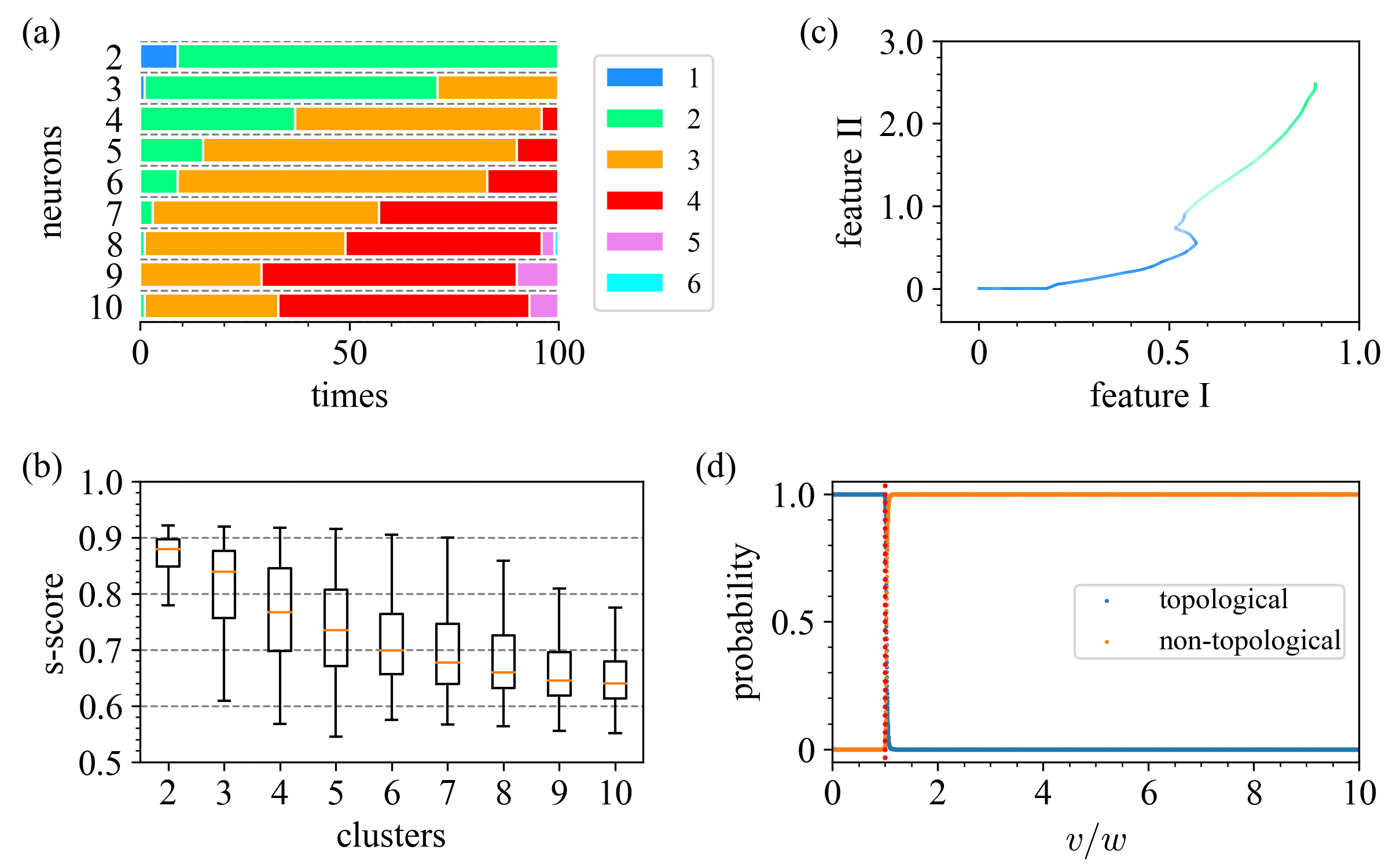}
		\caption{AE results for type-I data (MCMs) of SSH model. (a) The discrete distribution of necessary number of neurons $d_z$ for a given $n_{mid}$ neurons in the middlemost layer (2 to 10 along $y$-axis). Results from 100 independently trained AEs with type-I data are statistically calculated: The length of every color bar is proportional to the number of times that $d_z$ occurred within 100 models. Different color in the legend represents different $d_z$. (b) The box plot of the $s$-score as a function of $n$-clustering (via K-means method). (c) Latent representations projected to a subspace spanned by the first two principle components. Each color indicates its corresponding cluster (phase). (d) The neuron output ``phase diagram'' as a function of $v/w$ with $L=10$ for SSH model from a trained CNN by supervised learning in the last step of the ML pipeline.}
		\label{fig:AE_SSH_BA}
	\end{center}
\end{figure*}

Similar to the 1D $p$-wave superconductor case, we first train a neural network to encode our type-I input data to latent representations.  We determine the minimal dimension $d_z$ of the latent space by training a series of AEs with same model architecture \cite{appendix} except for the number of hidden neurons in the middlemost layer  ($n_{mid}$, from 2 to 10).  In Fig.~\ref{fig:AE_SSH_BA}(a), we show the discrete distribution of $d_z$, which indicates the necessary dimension of the latent representations to keep at least $99\%$ variance of them for each $n_{mid}$ by PCA, after repeating 100 times of the same training procedure. Clearly, $d_z$ is suggested to be 4.

Once $d_z$ is known, we take the latent representations of all input data from previously trained AEs with $n_{mid}=d_z$ and do SA to estimate the optimal number of clusters $n$ via K-means method. As shown in Fig.~\ref{fig:AE_SSH_BA}(b), it points out that the mean of Silhouette value reaches the highest one when $n=2$. In addition, by projecting the 4D latent representations into a 2D space spanned by the first two principle components (features), we have Fig.~\ref{fig:AE_SSH_BA}(c). By noticing the density change of the feature points, the plot suggests how the system could be consistently divided into two clusters (phases).

Due to the unconstrained nature of the latent representations in AEs, the phase transition point found is statistically at mean value $v=0.987$ with standard deviations 0.207, after collecting clustering results from 100 sets of latent representations via different trained AEs ($n_{mid}=4$). Although such critical value is very close to the theoretical value 1, it still gets non-ignorable deviation. To reduce this variance, we again train a CNN classifier \cite{appendix} in supervised learning manner to predict the whole phase diagram. We prepare a ``labeled'' dataset by picking up two seeds having highest Silhouette value in each cluster. They correspond to $v/w=0.332, 5.799$ of the original input ``images'', and for each point we expand symmetrically around its location by a window width $0.1w$ to obtain 2000 points with equal spacing. These 4000 data points are collected to be our training dataset, whereas the original 10,001 ones are formed our test set without labels. It turns out that the phase boundary predicted by trained CNN classifiers for the test set is clearly sharper at mean value 1.024 with much smaller deviation 0.041 (after repeating same training procedure 100 times), as shown in Fig.~\ref{fig:AE_SSH_BA}(d). 

\begin{figure}[tb]
	\begin{center}
		\includegraphics[width=7.5cm]{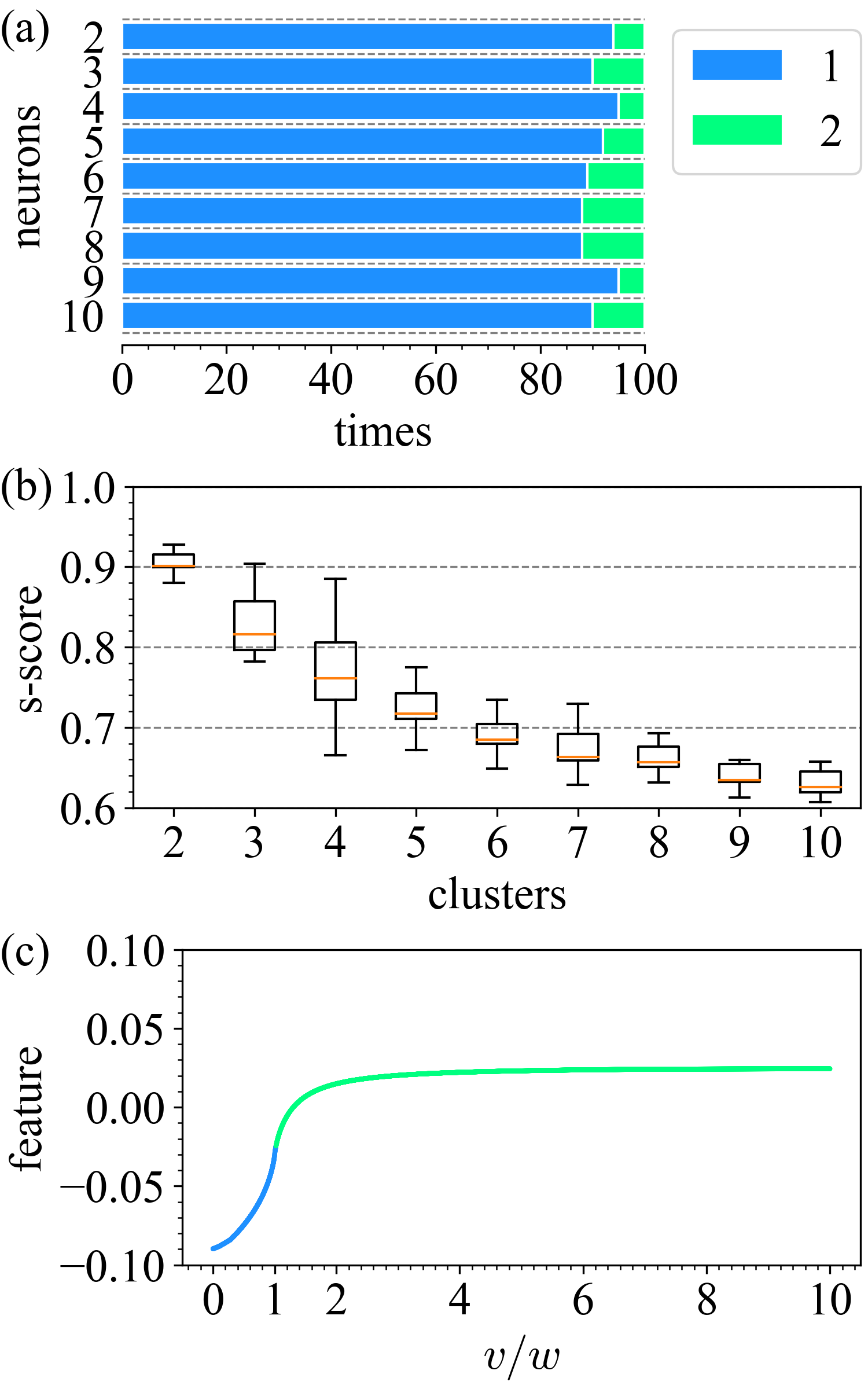}
		\caption{VAE results for type-I data (MCMs) of SSH model. (a) The discrete distribution of necessary number of neurons $d_z$ for a given $n_{mid}$ (paired) neurons in the middlemost layer (2 to 10 along $y$-axis). Results from 100 independently trained VAEs are statistically calculated: The dominant $d_z$ is clearly one (pair) here. (b) The box plot of the $s$-score as a function of $n$-clustering (via K-means method). (c) Latent representations (of type-I data) as a function of $v/w$. Each color indicates its corresponding cluster (phase).}
		\label{fig:VAE_SSH_BA}
	\end{center}
\end{figure}

\subsubsection{VAE approach}
Following the insights obtained from the 1D $p$-wave SC case, we employ VAE approach to impose some constraints on the encoded representations for a more stable solution. As a first step, we train a series of VAEs with $n_{mid}=2$ to 10 pairs of hidden neurons in the middlemost layer \cite{appendix}. Note that, similarly, one of the neurons in each pair outputs the mean of encoded representation distribution and the other produces its corresponding variance. After repeating the same training procedure 100 times for a given $n_{mid}$, the number of necessary dimension for the latent representations obtained after PCA to keep at least $99\%$ variance of them is simply one (pair), as clearly shown in Fig.~\ref{fig:VAE_SSH_BA}(a). This again proves the stability of obtaining robust latent representations via VAE method.

Since the minimal dimension of the encoded representations is one (pair), we then employs SA to estimate the best number of clusters $n$ for them via K-means method. Fig.~\ref{fig:VAE_SSH_BA}(b) clearly shows that the Silhouette value achieves the highest one when $n=2$. Moreover, Fig.~\ref{fig:VAE_SSH_BA}(c) depicts the encoded representations as a function of $v/w$ and different colors indicate distinct clusters (phases). Finally, from our repeated clustering results, the phase transition point is statistically found to be at mean value 1.051 with standard deviations 0.069. This deviation is relatively smaller than the one obtained from the AE approach. Therefore, the supervised learning in the last step in Fig.~\ref{fig:process}(a) is neglected here.

\section{Discussion and Conclusion}
The proposed ML procedure may has demonstrated its superiority, enough for recognizing phase transitions without prior or with rare knowledge on phases of matter, by taking advantages of unsupervised and (optionally) supervised learning algorithms. However, a few issues regarding with this approach are worth mentioning here.

(1) As commonly known, AE can make a non-linear dimensional reduction for the data, while PCA can only do a linear one. Thus, most importantly, one may ask whether AE or VAE is an essential component in the ML pipeline. To address this issue, we conduct more numerical experiments on the 1D $p$-wave SC with types I, II, and III data formats. For the most compressed data format, {\it i.e.}, type-II, we find that with or without AE (or VAE) plays no essential role on the later clustering. But this is not the case when considering type-I and type-III formats. Using solely PCA results often leads to consequences such as keeping higher necessary latent space dimension $d_z$ (to keep high variance) or getting wrong number of clusters (phases) via K-means method. The latter case could be related to the limitation of K-means method, which is notoriously known to fail for clustering concentric circles. So, one can view AE as an important component in order to consider general data formats.  

\begin{figure}[tb]
	\begin{center}
		\includegraphics[width=7.5cm]{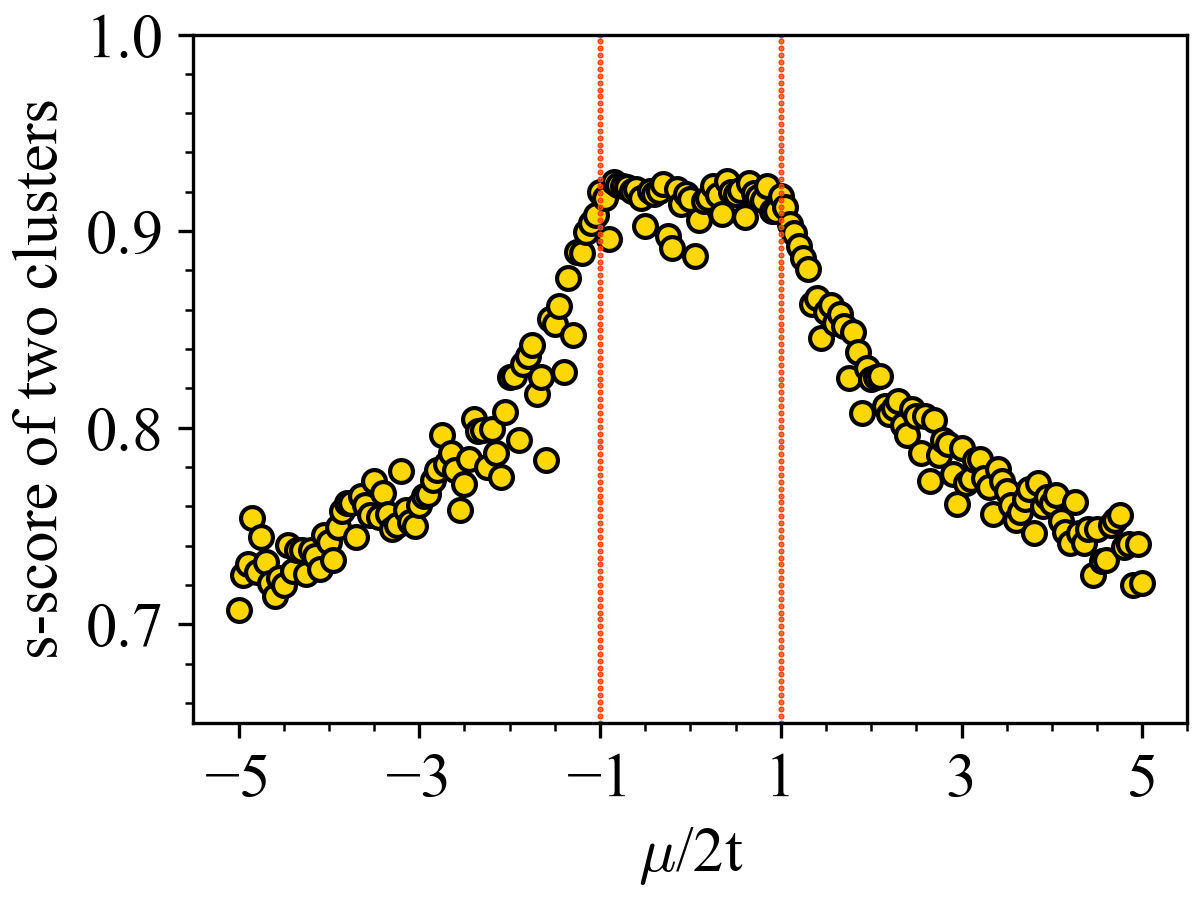}
		\caption{The $s$-score as a function of $\mu/2t$ for 1D $p$-SC at $L=10$. For a given $\mu/2t$, each yellow spot corresponds to an averaged $s$-score (over 10 trained VAEs) when enforcing the encoded input data of type-I, with $\Delta_/t$ varying from -10 to 10, to two clusters. Red lines indicate theoretical phase boundaries.}
		\label{fig:Delta}
	\end{center}
\end{figure}

(2) Another important aspect we haven't mentioned is the effect of varying $\Delta$, an essential piece for completing the whole phase diagram. As one can see in Fig.~\ref{fig:topological}(a), there are two phases in the region with $-1<\mu/2t<1$ when varying $\Delta$, while there is only one outside of it. This is a challenge for our proposed ML procedure because the K-means method and SA are not useful when number of clusters equals one. However, it is remarkable that our method still provides meaningful results. As shown in Fig.~\ref{fig:Delta}, each yellow spot corresponds to an averaged $s$-score when grouping the encoded input data into fixed two clusters in the 1D $p$-wave superconductor for a given $\mu/2t$. Note that the encoded data are generated from 10 VAEs with $n_{mid}=1$ for each $\mu$, trained by using type-I data format. Explicitly, they are collected from 20,001 equally-spaced data points in the range between $\Delta/t=-10$ and 10. Finally, one can clearly see that the averaged $s$-scores outside the region with $-1<\mu/2t<1$ are all suppressed, and this phenomenon implies the failure of our enforced clustering with $n=2$.  The decaying $s$-score indicates that $n$ should be 1 instead of 2.

\begin{figure}[t]
	\begin{center}
		\includegraphics[width=7.5cm]{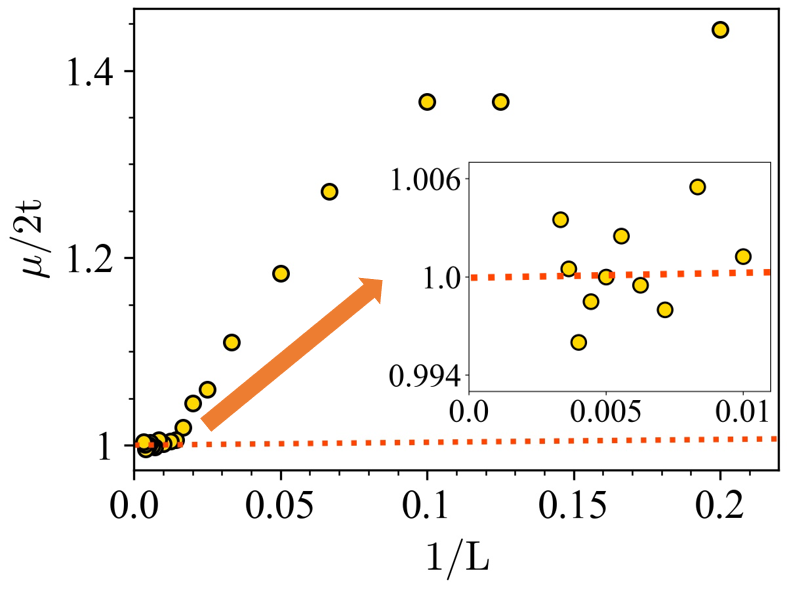}
		\caption{The critical phase transition points as a function of the subsystem size $L$ (up to 300) for 1D $p$-SCs at $\Delta/t=2$. Each yellow spot represents an averaged critical $\mu^*/2t$ based on 10 independently trained AEs for a given $L$. The red dashed line is a linear fit to data points of $L>100$, with interception 0.9995 on $y$-axis when extrapolating to $L=\infty$.}
		\label{fig:finite_size}
	\end{center}
\end{figure}

(3) To avoid some ``boundary effect'' that may degrade the validity of our proposed method, it should be additionally tested by considering the finite size effect. In particular, for a topological 1D $p$-wave SC the Majorana zero modes would appear at boundaries with coherent length proportional to $t/\Delta$, and thus are sensitive to the system size. For simplicity, let us take into account type-II data for a 1D $p$-wave SC at $\Delta/t=2$ and do not worry about fine tuning by supervised learning in the last step. The task here is to determine the phase boundary between topological and non-topological phases. As the size $L$ of the subsystem A increases from 5 to 300, Fig.~\ref{fig:finite_size} shows that the phase boundary converges to $\mu/2t = 0.9995$ by extrapolation to $L=\infty$. The value is very close to the theoretical one, namely, $\mu/2t=1$ and thus it further confirms our results in Sec. IV.

(4) Using neural networks for ML often raises a serious query about what the reasoning is behind model's predictions. This issue may degrade its value, in particular, for any scientific discovery. There is no exception in our proposal. However, to go one step further for the explainability, one may borrow some idea from feature selections in ML \cite{Phuong05,Ribeiro16,Shrikumar17,Lundberg21}. For instance, follow the proposed ML procedure and consider type-III input data, where each image is formed by eigenvectors of a BCM arranged in a certain way. If one masks the parts of all input images corresponding to the ``boundaries'' of the focused system, one can then observe its consequence after training. Our preliminary results show that the (phase) clustering becomes so poor that the information at boundaries is therefore not negligible. This also indirectly implies that the presence of edge modes or not helps model prediction. It, though not a complete solution, may shed some light on the opaque doubts of using DL in phase detection. 

In conclusion, we propose a ML procedure, mainly in an unsupervised learning manner, to study topological phase transitions in both 1D $p$-SC and SSH systems. This procedure includes a series of steps: feature extraction, dimensional reduction, clustering, Silhouette analysis, and fine tuning by supervised learning. Most importantly, three quantum-entanglement based quantities, MCMs, OPES and OPEEs (from BCMs), are considered to feed into neural networks for training. We find that in the feature extraction part VAEs provide more stable latent representations of the input data. Moreover, our results reliably reproduce the whole phase diagrams for both systems studied here, demonstrating the usefulness of our proposal without knowing prior knowledge of the phase space.  

\begin{acknowledgments}
M.C.Chung acknowledges the MoST support under the contract NO. 108-2112-M-005 -010 -MY3 and Asian Office of Aerospace Research and Development (AOARD) for the support under the award NO. FA2386-20-1-4049. 
\end{acknowledgments}

\appendix
\section{Model architectures}
The basic structures for both AEs and VAEs are already schematically shown in Figs.~\ref{fig:process}(b) and (c). Here we present explicit model details for our numerical results shown in all figures. 

In the case of 1D $p$-wave SCs, for the AE used for Fig.~\ref{fig:AE_1DSC_BA} the model architecture is given in Table~\ref{tab:AE_SC},

\begin{table}[htbp]
	\centering
	\renewcommand\arraystretch{1.3} 
	\begin{tabular}{|p{2.5cm}<{\centering}|p{1.5cm}<{\centering}|p{1.6cm}<{\centering}|p{1.8cm}<{\centering}|}
			\hline
			\multicolumn{4}{|c|}{Fig. 4 (AE)} \\ 
			\hline
			Layer & Params & Activation & Batch norm \\
			\hline
			\multicolumn{4}{|c|}{Input:10$\times$10$\times$1} \\
			\hline
			Conv & 3$\times$3$\times$8 & ReLU & True \\
			\hline
			Residual block & \multicolumn{3}{c|}{output:10$\times$10$\times$8} \\
			\hline
			Average pooling & 2$\times$2 & & \\
			\hline
			Linear & 200$\times$256 & Tanh & True \\
			\hline
			Linear & 256$\times$128 & Tanh & True \\
			\hline
			Linear & 128$\times$32 & Tanh & True \\
			\hline
			Linear & 32$\times$8 & Tanh & True \\
			\hline
			Linear & 8$\times$4 & Tanh & True \\
			\hline
			Linear & 4$\times$8 & Tanh & True \\
			\hline
			Linear & 8$\times$32 & Tanh & True \\
			\hline
			Linear & 32$\times$128 & Tanh & True \\
			\hline
			Linear & 128$\times$256 & Tanh & True \\
			\hline
			Linear & 256$\times$200 & Tanh & True \\
			\hline
			Up Sampling & 2$\times$2 &   &  \\
			\hline
			Transposed Conv & 3$\times$3$\times$8 &  & \\
			\hline
			Transposed Conv & 3$\times$3$\times$8 &  & \\
			\hline
			Transposed Conv & 3$\times$3$\times$8 &  & \\
			\hline
			Transposed Conv & 3$\times$3$\times$8 &  & \\
			\hline
			Transposed Conv & 3$\times$3$\times$1 & Sigmoid & \\
			\hline
	\end{tabular}
    \caption{Model architecture of AE used for Fig.~\ref{fig:AE_1DSC_BA}.}
	\label{tab:AE_SC}
\end{table}
\newpage

where the content of a residual block \cite{He15} is separately shown in Table~\ref{tab:RB_SC}.

\begin{table}[htbp]
	\centering
	\renewcommand\arraystretch{1.3} 
	\begin{tabular}{|p{1.5cm}<{\centering}|p{2.5cm}<{\centering}|p{1.6cm}<{\centering}|p{1.8cm}<{\centering}|}
		\hline
		\multicolumn{4}{|c|}{Residual Block} \\ 
		\hline
		Layer & Params & Activation & Batch norm \\ 
		\hline
		\multicolumn{4}{|l|}{Left:} \\
		\hline
		Conv & 3$\times$3$\times$outchannel & ReLU & True \\ 
		\hline
		Conv & 3$\times$3$\times$outchannel &   & True \\ 
		\hline
		\multicolumn{4}{|l|}{Shortcut:} \\
		\hline
		Conv & 1$\times$1$\times$outchannel &   & True \\ 
		\hline
		\multicolumn{4}{|c|}{Left+Shortcut} \\
		\hline
		\multicolumn{4}{|c|}{ReLU} \\
		\hline
	\end{tabular}
    \caption{The residual block details in AE.}
    \label{tab:RB_SC}
\end{table}

Moreover, as to fine tuning the phase boundaries, a CNN model is employed and shown in Table~\ref{tab:CNN_SC}.

\begin{table}[htbp]
	\centering
	\renewcommand\arraystretch{1.3} 
	\begin{tabular}{|p{2.5cm}<{\centering}|p{1.5cm}<{\centering}|p{1.6cm}<{\centering}|p{1.8cm}<{\centering}|}
		\hline
		\multicolumn{4}{|c|}{Fig. 4 (CNN)} \\ 
		\hline
		Layer & Params & Activation & Batch norm \\
		\hline
		\multicolumn{4}{|c|}{Input:10$\times$10$\times$1} \\
		\hline
		Conv & 3$\times$3$\times$16 & ReLU & True \\
		\hline
		Residual block & \multicolumn{3}{c|}{output:10$\times$10$\times$32} \\
		\hline
		Average pooling & 4$\times$4 & & \\
		\hline
		Linear & 128$\times$16 & ReLU &  \\
		\hline
		Linear & 16$\times$3 & ReLU &  \\
		\hline
		Linear & 3 & Softmax &  \\
		\hline
	\end{tabular}
	\caption{The CNN model used for Fig.~\ref{fig:AE_1DSC_BA}(d).}	
	\label{tab:CNN_SC}
\end{table}

s to the AE used in Fig.~\ref{fig:AE_1DSC_BCMev}, the model architecture is given in Table~\ref{tab:AE_SC_II}.

\begin{table}[htbp]
	\centering
	\renewcommand\arraystretch{1.3}  
	\begin{tabular}{|p{2.5cm}<{\centering}|p{2.5cm}<{\centering}|p{2.5cm}<{\centering}|}
		\hline
		\multicolumn{3}{|c|}{Fig. 5 (AE)} \\ 
		\hline
		\multicolumn{3}{|c|}{Input:20$\times$1} \\
		\hline
		Layer & Params & Activation \\
		\hline
		Linear & 20$\times$16 & ReLU \\
		\hline
		Linear & 16$\times$2 & ReLU \\
		\hline
		Linear & 2$\times$16 & ReLU \\
		\hline
		Linear & 16$\times$20 & Sigmoid \\
		\hline
	\end{tabular}
	\caption{Model architecture of AE used for Fig.~\ref{fig:AE_1DSC_BCMev}.}
	\label{tab:AE_SC_II}
\end{table}
\newpage

For the AE used in Fig.~\ref{fig:AE_1DSC_BCMevec}, the model architecture is given in Table~\ref{tab:AE_SC_III}.

\begin{table}[htbp]
	\centering
	\renewcommand\arraystretch{1.3} 
	\begin{tabular}{|p{3cm}<{\centering}|p{1.6cm}<{\centering}|p{1.6cm}<{\centering}|p{1.8cm}<{\centering}|}
		\hline
		\multicolumn{4}{|c|}{Fig. 6 (AE)} \\ 
		\hline
		Layer & Params & Activation & Batch norm \\
		\hline
		\multicolumn{4}{|c|}{Input:20$\times$20$\times$1} \\
		\hline
		Conv & 20$\times$20$\times$4 & ReLU & True \\
		\hline
		Residual block & \multicolumn{3}{c|}{output:20$\times$20$\times$64} \\
		\hline
		Residual block & \multicolumn{3}{c|}{output:20$\times$20$\times$128} \\
		\hline
		GlobAverage pooling & 128 & & \\
		\hline
		Linear & 128$\times$32 & ReLU &  \\
		\hline
		Linear & 32$\times$8 & ReLU &  \\
		\hline
		Linear & 8$\times$32 & ReLU &  \\
		\hline
		Linear & 32$\times$128 & ReLU &  \\
		\hline
		Linear & 128$\times$12800 &  &  \\
		\hline
		Up Sampling & 2$\times$2 &  &  \\
		\hline
		Transposed Conv & 3$\times$3$\times$128 &  & \\
		\hline
		Transposed Conv & 3$\times$3$\times$64 &  & \\
		\hline
		Transposed Conv & 3$\times$3$\times$64 &  & \\
		\hline
		Transposed Conv & 5$\times$5$\times$4 &  & \\
		\hline
		Transposed Conv & 5$\times$5$\times$1 & Sigmoid & \\
		\hline
	\end{tabular}
	\caption{Model architecture of AE used for Fig.~\ref{fig:AE_1DSC_BCMevec}.}
	\label{tab:AE_SC_III}
\end{table}

Similarly, a CNN model is employed to fine tune the phase boundaries and shown in Table~\ref{tab:CNN_SC_III}.

\begin{table}[htbp]
	\centering
	\renewcommand\arraystretch{1.3} 
	\begin{tabular}{|p{3cm}<{\centering}|p{1.5cm}<{\centering}|p{1.6cm}<{\centering}|p{1.8cm}<{\centering}|}
		\hline
		\multicolumn{4}{|c|}{Fig. 6 (CNN)} \\ 
		\hline
		Layer & Params & Activation & Batch norm \\
		\hline
		\multicolumn{4}{|c|}{Input:200$\times$200$\times$1} \\
		\hline
		Residual block & \multicolumn{3}{c|}{output:200$\times$200$\times$32} \\
		\hline
		Residual block & \multicolumn{3}{c|}{output:200$\times$200$\times$64} \\
		\hline
		Residual block & \multicolumn{3}{c|}{output:200$\times$200$\times$128} \\
		\hline
		GlobAverage pooling & 128 & & \\
		\hline
		Linear & 128$\times$64 & ReLU &  \\
		\hline
		Linear & 64$\times$3 & ReLU &  \\
		\hline
		Linear & 3 & Softmax &  \\
		\hline
	\end{tabular}
	\caption{The CNN model used for Fig.~\ref{fig:AE_1DSC_BCMevec}(d).}
	\label{tab:CNN_SC_III}
\end{table}
\newpage

On the other hand, for the VAE used for Fig.~\ref{fig:VAE_1DSC_BA}, the model architecture is given in Table~\ref{tab:VAE_SC}.
\begin{table}[htbp]
	\centering
	\renewcommand\arraystretch{1.3}  
	\begin{tabular}{|p{2.5cm}<{\centering}|p{2.5cm}<{\centering}|p{2.5cm}<{\centering}|}
		\hline
		\multicolumn{3}{|c|}{Fig. 8 , Fig. 12 , Fig. 13 (VAE)} \\ 
		\hline
		Layer & Params & Activation \\
		\hline
		\multicolumn{3}{|c|}{Input:10$\times$10$\times$1} \\
		\hline
		Linear & 100$\times$128 & ReLU \\
		\hline
		Linear & 128$\times$64 & ReLU \\
		\hline
		Linear & 64$\times$32 & ReLU \\
		\hline
		Linear & 32$\times$16 & ReLU \\
		\hline
		Linear & 16$\times$1 & ReLU \\
		\hline
		Linear & 1$\times$16 & ReLU \\
		\hline
		Linear & 16$\times$32 & ReLU \\
		\hline
		Linear & 32$\times$64 & ReLU \\
		\hline
		Linear & 64$\times$128 & ReLU \\
		\hline
		Linear & 128$\times$100 & Sigmoid \\
		\hline
	\end{tabular}
	\caption{The CNN model used for Figs.~\ref{fig:VAE_1DSC_BA}, \ref{fig:VAE_SSH_BA}, and \ref{fig:Delta}.}
	\label{tab:VAE_SC}
\end{table}

As to the VAE used in Fig.~\ref{fig:VAE_1DSC_BCMev}, the model architecture is given in Table~\ref{tab:VAE_SC_II}.
\begin{table}[htbp]
	\centering
	\renewcommand\arraystretch{1.3}  
	\begin{tabular}{|p{2.5cm}<{\centering}|p{2.5cm}<{\centering}|p{2.5cm}<{\centering}|}
		\hline
		\multicolumn{3}{|c|}{Fig. 9 (VAE)} \\
		\hline
		\multicolumn{3}{|c|}{Input:20$\times$1} \\ 
		\hline
		Layer & Params & Activation \\
		\hline
		Linear & 20$\times$128 & ReLU \\
		\hline
		Linear & 128$\times$64 & ReLU \\
		\hline
		Linear & 64$\times$32 & ReLU \\
		\hline
		Linear & 32$\times$16 & ReLU \\
		\hline
		Linear & 16$\times$1 & ReLU \\
		\hline
		Linear & 1$\times$16 & ReLU \\
		\hline
		Linear & 16$\times$32 & ReLU \\
		\hline
		Linear & 32$\times$64 & ReLU \\
		\hline
		Linear & 64$\times$128 & ReLU \\
		\hline
		Linear & 128$\times$20 & Sigmoid \\
		\hline
	\end{tabular}
	\caption{The VAE model used for Fig.~\ref{fig:VAE_1DSC_BCMev}.}
	\label{tab:VAE_SC_II}
\end{table}
\newpage

Finally, for the VAE used in Fig.~\ref{fig:VAE_1DSC_BCMevec}, the model architecture is given in Table~\ref{tab:VAE_SC_III}.
\begin{table}[htbp]
	\centering
	\renewcommand\arraystretch{1.3}  
	\begin{tabular}{|p{2.5cm}<{\centering}|p{2.5cm}<{\centering}|p{2.5cm}<{\centering}|}
		\hline
		\multicolumn{3}{|c|}{Fig. 10 (VAE)} \\ 
		\hline
		Layer & Params & Activation \\
		\hline
		\multicolumn{3}{|c|}{Input:20$\times$20$\times$1} \\ 
		\hline
		Linear & 400$\times$256 & ReLU \\
		\hline
		Linear & 256$\times$196 & ReLU \\
		\hline
		Linear & 196$\times$128 & ReLU \\
		\hline
		Linear & 128$\times$96 & ReLU \\
		\hline
		Linear & 96$\times$32 & ReLU \\
		\hline
		Linear & 32$\times$1 & ReLU \\
		\hline
		Linear & 1$\times$32 & ReLU \\
		\hline
		Linear & 32$\times$96 & ReLU \\
		\hline
		Linear & 96$\times$128 & ReLU \\
		\hline
		Linear & 128$\times$196 & ReLU \\
		\hline
		Linear & 196$\times$256 & ReLU \\
		\hline
		Linear & 256$\times$400 & Sigmoid \\
		\hline
	\end{tabular}
	\caption{The VAE model used for Fig.~\ref{fig:VAE_1DSC_BCMevec}.}
	\label{tab:VAE_SC_III}
\end{table}

In the case of SSH models, for the AE used for Fig.~\ref{fig:AE_SSH_BA} the model architecture is given in Table~\ref{tab:AE_SSH},
\begin{table}[htbp]
	\centering
	\renewcommand\arraystretch{1.3} 
	\begin{tabular}{|p{2.5cm}<{\centering}|p{1.5cm}<{\centering}|p{1.6cm}<{\centering}|p{1.8cm}<{\centering}|}
		\hline
		\multicolumn{4}{|c|}{Fig. 11 (AE)} \\ 
		\hline
		Layer & Params & Activation & Batch norm \\
		\hline
		\multicolumn{4}{|c|}{Input:10$\times$10$\times$1} \\ 
		\hline
		Conv & 10$\times$10$\times$4 & ReLU & True \\
		\hline
		Residual block & \multicolumn{3}{c|}{output:10$\times$10$\times$4} \\
		\hline
		Average pooling & 2$\times$2 & & \\
		\hline
		Linear & 100$\times$64 & ReLU & True \\
		\hline
		Linear & 64$\times$16 & ReLU & True \\
		\hline
		Linear & 16$\times$4 & ReLU & True \\
		\hline
		Linear & 4$\times$16 & ReLU & True \\
		\hline
		Linear & 16$\times$64 & ReLU & True \\
		\hline
		Linear & 64$\times$100 &  & True \\
		\hline
		Up Sampling & 2$\times$2  &   &  \\
		\hline
		Transposed Conv & 3$\times$3$\times$4 &  & \\
		\hline
		Transposed Conv & 3$\times$3$\times$4 &  & \\
		\hline
		Transposed Conv & 3$\times$3$\times$4 &  & \\
		\hline
		Transposed Conv & 3$\times$3$\times$4 &  & \\
		\hline
		Transposed Conv & 3$\times$3$\times$1 & Sigmoid & \\
		\hline
	\end{tabular}
	\caption{The AE model used for Fig.~\ref{fig:AE_SSH_BA}.}
	\label{tab:AE_SSH}
\end{table}
\newpage

where the residual block is again shown in Table~\ref{tab:RB_SC}.  A CNN model is employed to fine tune the phase boundaries then and is shown in Table~\ref{tab:CNN_SSH_BA}.
\begin{table}[htbp]
	\centering
	\renewcommand\arraystretch{1.3} 
	\begin{tabular}{|p{2.5cm}<{\centering}|p{1.5cm}<{\centering}|p{1.6cm}<{\centering}|p{1.8cm}<{\centering}|}
		\hline
		\multicolumn{4}{|c|}{Fig. 11 (CNN)} \\ 
		\hline
		Layer & Params & Activation & Batch norm \\
		\hline
		\multicolumn{4}{|c|}{Input:10$\times$10$\times$1} \\ 
		\hline
		Conv & 3$\times$3$\times$16 & ReLU & True \\
		\hline
		Residual block & \multicolumn{3}{c|}{output:10$\times$10$\times$32} \\
		\hline
		Average pooling & 4$\times$4 & & \\
		\hline
		Linear & 128$\times$16 & ReLU &  \\
		\hline
		Linear & 16$\times$3 & Softmax &  \\
		\hline
	\end{tabular}
	\caption{The CNN model used for Fig.~\ref{fig:AE_SSH_BA}(d).}
	\label{tab:CNN_SSH_BA}
\end{table}

Furthermore, for the VAE used for Fig.~\ref{fig:VAE_SSH_BA}, the model architecture is given in Table~\ref{tab:VAE_SC}.

The VAE model used for examining the effect of varying $\Delta$ and the AE model used for the finite-size study on 1D $p$-wave SCs are given in Table~\ref{tab:VAE_SC} and Table~\ref{tab:FS}, respectively.
\begin{table}[htbp]
	\centering
	\renewcommand\arraystretch{1.3}  
	\begin{tabular}{|p{2.5cm}<{\centering}|p{2.5cm}<{\centering}|p{2.5cm}<{\centering}|}
		\hline
		\multicolumn{3}{|c|}{Fig. 14 (AE)} \\ 
		\hline
		Layer & Params & Activation \\
		\hline
		\multicolumn{3}{|c|}{Input:2$\times$L$\times$1} \\
		\hline
		Linear & 2$\times$L$\times$128 & ReLU \\
		\hline
		Linear & 128$\times$64 & ReLU \\
		\hline
		Linear & 64$\times$32 & ReLU \\
		\hline
		Linear & 32$\times$16 & ReLU \\
		\hline
		Linear & 16$\times$1 & ReLU \\
		\hline
		Linear & 1$\times$16 & ReLU \\
		\hline
		Linear & 16$\times$32 & ReLU \\
		\hline
		Linear & 32$\times$64 & ReLU \\
		\hline
		Linear & 64$\times$128 & ReLU \\
		\hline
		Linear & 128$\times$2$\times$L & Sigmoid  \\
		\hline
	\end{tabular}
	\caption{The AE model used for Fig.~\ref{fig:finite_size}.}
	\label{tab:FS}
\end{table}


\begin{thebibliography}{30}
\bibitem{ssh_topological} In fact, if one slightly breaks time-reversal and particle-hole symmetries of a SSH model, it would become a topological system characterized by a $Z_2$ number within AIII class.
\bibitem{prescription} $99\%$ is simply an empirical value to help select out an optimal dimension for a feature space without keeping too much noisy information.
\bibitem{typeiii} The input training data (made of OPEEs) taken here are not squared in order to follow the same supervised learning procedure adopted in the type-I (MCM) case for refining the phase boundaries.
\end{thebibliography}
\end{document}